\documentclass[11pt]{article}
\usepackage{amssymb,amsmath}
\usepackage{graphicx}
\usepackage{cite}
\usepackage{xcolor}
\definecolor{darkblue}{rgb}{0,0,0.8}
\usepackage[dvipdfmx,breaklinks,colorlinks=true,linktoc=page,urlcolor=darkblue,linkcolor=darkblue,citecolor=darkblue]{hyperref}
\usepackage{overpic}
\usepackage{simplewick}
\usepackage{url}

\numberwithin{equation}{section}
\newcommand{\bG}{\mathbf{G}}
\newcommand{\bS}{\mathbf{S}}
\newcommand{\rmi}{\mathrm{i}}
\newcommand{\rme}{\mathrm{e}}
\newcommand{\mini}[1]{{\scalebox{0.6}{$#1$}}}

\begin{document}

\thispagestyle{empty}

\vskip1cm
\begin{center}
{\Large \bf Degenerate chords in double-scaled SYK}
\vskip1.5cm
\renewcommand{\thefootnote}{\fnsymbol{footnote}}
Kazuo Hosomichi$^{1,2}$\footnote{hosomiti@nda.ac.jp} and
Masayoshi Sato$^1$\footnote{ed23003@nda.ac.jp}

\bigskip
{\footnotesize
$^1$\it 
Department of Applied Physics, School of Applied Sciences,
National Defense Academy,\\
1-10-20 Hashirimizu, Yokosuka-city, Kanagawa 239-8686 Japan \\
$^2$\it 
Human Resources Development Division, Bureau of Personnel and Education,
Japan Ministry of Defense,\\
5-1 Honmura-cho, Ichigaya, Shinjuku-guy, Tokyo 162-8801 Japan\\
}
\end{center}

\vskip1cm
\begin{abstract}
\noindent
The matter operator in the double-scaled SYK model exhibits special properties when its dimension is analytically continued to $-\frac12$. At this dimension, the operator is in a degenerate representation of the $q$-deformed oscillator algebra and satisfies a null vector equation. Its peculiar fusion property gives rise to recursion relations among matter correlation functions. We find that these relations allow us to determine the two-point function without having to sum over infinitely many chord diagrams.

\end{abstract}
\newpage
\setcounter{tocdepth}{2}
\tableofcontents
%%%%%%%%%%%%%%%%%%%%%%%%%%
\renewcommand{\thefootnote}{\arabic{footnote}}
\setcounter{footnote}{0}

\section{Introduction}\label{sec:intro}

The Sachdev-Ye-Kitaev (SYK) model \cite{Sachdev:1992fk,Kitaev:2015ta} is a quantum mechanical model of $N$ Majorana fermions with a Hamiltonian which is a degree-$p$ homogeneous polynomial in the fermions. This model is known to be highly chaotic, and has maximal chaos exponent. It is also known that, after averaging over the random coupling, the theory can be described in terms of $\mathrm{O}(N)$-singlet bilocal fields \cite{Maldacena:2016hyu}. The IR physics is governed by an emergent 1d reparametrization invariance which is both spontaneously and explicitly broken. The theory is thus described effectively by the 1d Schwarzian action, which is also the effective action of the Jackiw-Teitelboim theory \cite{Jackiw:1984je,Teitelboim:1983ux} of $\text{AdS}_2$ gravity coupled to a dilaton field.

The SYK model have been used to study various problems in quantum gravity from the perspective of AdS/CFT duality\cite{Almheiri:2014cka,Maldacena:2016upp,Jensen:2016pah,Engelsoy:2016xyb}. As an example, in a seminal work \cite{Cotler:2016fpe} it was found that the spectral form factor of the model does not decay permanently at late times, which is indicative of the discreteness of the black hole microstate spectrum. It was also found there that the late time behavior of the model is very similar to that of random matrices. Another more rigorous relationship to random matrix theory was discovered in \cite{Saad:2019lba}. It was shown there that the partition functions of JT gravity on surfaces with different number of handles and holes correspond to the topological expansion of a certain double-scaled matrix integral.

In this paper we focus on the SYK model in the double-scaling (DS) limit \cite{Erdos:2014zgc,Cotler:2016fpe} which sends $N,p\to\infty$ with $\lambda\equiv \frac{2p^2}N$ fixed. In this limit, the computation of observables reduces to summing over the so-called chord diagrams \cite{Berkooz:2018qkz,Berkooz:2018jqr,Berkooz:2024lgq}, which consist of a disk with an even number of dots along the boundary circle that are pairwise connected by chords. Furthermore, by introducing the chord Hilbert space, the same problem can be reformulated as a quantum mechanics of a $q$-deformed oscillator in which the creation and annihilation operators ${\sf a}^\dagger, {\sf a}$ satisfy \cite{Arik:1973vg}
\[
 {\sf a}{\sf a}^{\!\dagger} - q{\sf a}^{\!\dagger}{\sf a}\;=\; 1.\qquad
 \left(q\equiv\rme^{-\lambda}\right)
\]
Using these techniques, in \cite{Berkooz:2018qkz,Berkooz:2018jqr} the two- and four-point functions of matter operators were computed exactly and a set of diagrammatic rules for general correlation functions was proposed. The results obtained there indicate that the DSSYK model possesses an underlying ${\cal U}_{q^{1/2}}(\mathfrak{su}_{1,1})$ quantum group structure.

Chord diagrams also provide a natural link between quantities in the DSSYK model and bulk 2d gravity. For example, the number eigenstate $|n\rangle$ of the $q$-deformed oscillator algebra was shown to correspond to a state in the Hilbert space of bulk gravity of length $\ell=\lambda n$ \cite{Lin:2022rbf,Okuyama:2022szh}. The correspondence between the Hilbert spaces ${\cal H}_\text{SYK}\leftrightarrow{\cal H}_\text{bulk}$ has been studied in a more systematic manner using the two-sided Hilbert space formalism \cite{Lin:2023trc,Okuyama:2024yya,Okuyama:2024gsn}, in which a chord diagram is regarded as an inner product of two half-disks each defining a quantum state of a two-sided bulk geometry. A realization of the quantum group symmetry acting on the two-sided Hilbert space was recently found in \cite{vanderHeijden:2025zkr}. There have been many proposals to understand better the discretization of bulk spacetime and the associated mathematical structure using non-commutative $\mathrm{AdS}_2$ \cite{Berkooz:2022mfk,Almheiri:2024ayc,Berkooz:2024lgq}, particles on a quantum group manifold, sine-dilaton gravity \cite{Blommaert:2023opb,Blommaert:2024ymv,Blommaert:2024whf,Blommaert:2025avl,Belaey:2025ijg} and complex Liouville string theory \cite{Blommaert:2025eps}.

Despite the powerfulness of the chord diagrammatics, the actual analysis of the DSSYK model often requires some highly specialized knowledge of special functions such as $q$-hypergeometric series. The goal of this paper is to develop an alternative method that can reproduce the known exact results without such expertise, and use it to explore new aspects of the model. Our idea is to make use of the algebraic structure that shows up in the matrix elements and the fusion property of matter operators. We will especially focus on the matter operator $M_\mini{\Delta}$ of dimension $\Delta=-\frac12$ which has a number of interesting special properties. Our approach is similar in spirit to that of \cite{Fateev:2000ik} in which some fundamental structure functions of boundary Liouville theory was obtained by combining the field-theoretic and representation-theoretic analyses.

\paragraph{Organization of the paper}

In Section \ref{sec:DSSYK} we review the basic techniques to study the DSSYK model such as chord diagrams and $q$-deformed oscillator, and summarize some exact formulae for correlation functions. In Section \ref{sec:BLCFT} we give a brief introduction to the boundary Liouville CFT, highlighting the special properties of the boundary operator $B_\beta$ with momentum $\beta=-\frac b2$ (where $b$ is the Liouville coupling) and explaining how one can use them to derive the disk two-point function of boundary operators. In Section \ref{sec:deg} we study the properties of the operator ${\sf D}\equiv {\sf M}_\mini{-\frac12}$ in the DSSYK model which behaves in a very similar manner as $B_{-\frac b2}$ in Liouville CFT. First of all, we point out that its matrix element $\langle\theta_1|{\sf D}|\theta_2\rangle$ between the eigenstates of the chord Hamiltonian ${\sf H}={\sf a}+{\sf a}^\dagger$ has a delta-functional support. We argue that it follows from the null vector equation:
\[
 \chi({\sf D})\;\equiv\; {\sf H}^2{\sf D}-(q^{\frac12}+q^{-\frac12}){\sf HDH}+{\sf D}{\sf H}^2+(q^{-1}-1){\sf D}\;=\;0.
\]
We then study the operator product ${\sf M}_\mini{\Delta}\times{\sf M}_\mini{\Delta'}$ using the two-sided Hilbert space formalism, and find a special behavior when $\Delta=-\frac12$. We also obtain an explicit formula which expresses ${\sf M}_\mini{\Delta\mp\frac12}$ as composites of ${\sf D}, {\sf M}_\mini{\Delta}$ and ${\sf H}$. Furthermore, by studying the relation between the two-sided Hilbert spaces before and after the OPE ${\sf D}\times{\sf M}_\mini{\Delta}\to\sum_\pm{\sf M}_\mini{\Delta\mp\frac12}$, we derive a set of recursion relations which can reproduce the known matter two-point function. Finally, in Section \ref{sec:concl} we conclude with discussions on possible future directions.

\section{SYK model}\label{sec:DSSYK}

The SYK model is a quantum mechanical model of $N$ Majorana fermions $\psi_i~(i=1,\cdots,N)$ obeying $\{\psi_i,\psi_j\}=2\delta_{ij}$. The Hamiltonian is given by
\begin{equation}
H\;=\;\rmi^{p/2}\sum_{1\le i_1<\cdots<i_p\le N}J_{i_1\cdots i_p}\psi_{i_1}\cdots\psi_{i_p}.
\label{defH}
\end{equation}
We take the {\it disorder average} over the theory with different values of coupling $J_{i_1\cdots i_p}$ assuming that they obey Gaussian distribution with zero mean and
\begin{equation}
\langle J_{i_1\cdots i_p}J_{j_1\cdots j_p}\rangle_J =
\frac{{\cal J}^2}{\big({}^N_{\,p}\big)}\delta_{i_1j_1}\cdots\delta_{i_pj_p}.
\label{avJ} 
\end{equation}
Hereafter we set the dimensionful coupling ${\cal J}=1$.

\paragraph{DS limit and chord diagrams}

In the double-scaling limit \cite{Cotler:2016fpe, Erdos:2014zgc}
\begin{equation}
N\to\infty,\quad
p\to\infty\quad\text{with}\quad
\frac{2p^2}N=\lambda\quad\text{fixed},
\label{ds}
\end{equation}
the computation of various observables reduces to summation over the so-called chord diagrams \cite{Erdos:2014zgc,Berkooz:2018qkz,Berkooz:2018jqr}. As the most basic example, let us review here the evaluation of thermal partition function
\begin{align}
Z(\beta) &\;=\; \left\langle\mathrm{Tr}(\rme^{-\beta H})\right\rangle_J\;=\;\sum_{n=0}^\infty\frac{(-\beta)^n}{n!}m_n,\nonumber \\
m_n&\;=\; \left\langle\mathrm{Tr}H^n\right\rangle_J \;=\;\rmi^{\frac{np}2}\sum_{I_1,\cdots,I_n}\langle J_{I_1}\cdots J_{I_n}\rangle_J\;\mathrm{Tr}(\psi_{I_1}\cdots\psi_{I_n}).
\label{Zbeta}
\end{align}
Here $I_1,\cdots,I_n$ denote $p$-index sets and $\psi_I\equiv \psi_{i_1}\cdots\psi_{i_p}$ for $I=\{i_1,\cdots,i_p\}$. Note also that we normalize the trace so that $\mathrm{Tr}(1)=1$.

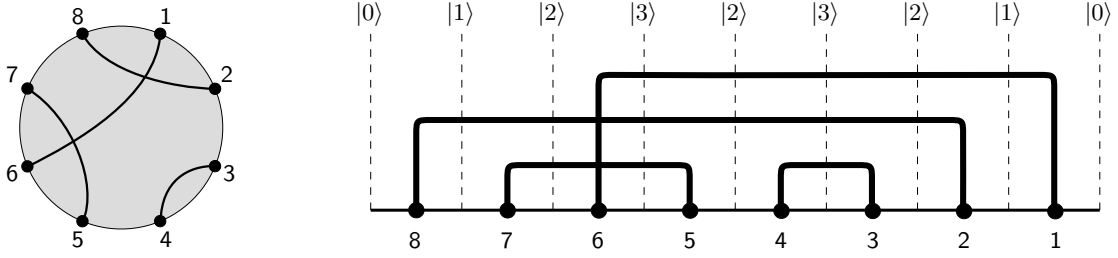
\begin{figure}[t]
\begin{center}
\begin{tabular}{ccc}
\begin{overpic}[height=30mm]{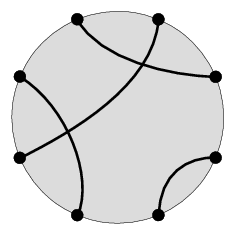}
\put(67,96){\sffamily\footnotesize 1}
\put(94,70){\sffamily\footnotesize 2}
\put(95,26){\sffamily\footnotesize 3}
\put(67,-3){\sffamily\footnotesize 4}
\put(28,-3){\sffamily\footnotesize 5}
\put(-1,26){\sffamily\footnotesize 6}
\put(-1,70){\sffamily\footnotesize 7}
\put(28,96){\sffamily\footnotesize 8}
\end{overpic}
&\qquad\qquad&
\begin{overpic}[height=32mm]{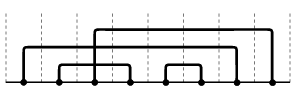}
\put(96,29){\footnotesize $|0\rangle$}
\put(84,29){\footnotesize $|1\rangle$}
\put(72,29){\footnotesize $|2\rangle$}
\put(60,29){\footnotesize $|3\rangle$}
\put(48,29){\footnotesize $|2\rangle$}
\put(36,29){\footnotesize $|3\rangle$}
\put(24,29){\footnotesize $|2\rangle$}
\put(12,29){\footnotesize $|1\rangle$}
\put( 0,29){\footnotesize $|0\rangle$}
\put(91.4,-1){\sffamily\footnotesize 1}
\put(79.3,-1){\sffamily\footnotesize 2}
\put(67.3,-1){\sffamily\footnotesize 3}
\put(55.2,-1){\sffamily\footnotesize 4}
\put(43.2,-1){\sffamily\footnotesize 5}
\put(31.1,-1){\sffamily\footnotesize 6}
\put(19.1,-1){\sffamily\footnotesize 7}
\put( 7,-1){\sffamily\footnotesize 8}
\end{overpic}
\end{tabular}
\end{center}
\caption{(left) a chord diagram contributing $q^2$ to $m_8$, and (right) a quantum mechanics interpretation of the same diagram.}
\label{fig:chord}
\end{figure}

After using Wick's theorem to evaluate $\langle\cdots \rangle_J$, the $n$ operators $\psi_{I_1},\cdots,\psi_{I_n}$ inside the trace in (\ref{Zbeta}) form $\frac n2$ pairs having the same $p$-index sets, and one is left with the summation over different pairings. Note that each pair can take $\big({}^N_{\,p}\big)$ different index sets, and it cancels with the factor in the denominator of (\ref{avJ}). The trace $\mathrm{Tr}(\psi_{I_1}\cdots\psi_{I_n})$ for a given pairing is then evaluated by permuting the $n$ operators until the paired operators sit next to each other and using $\rmi^p(\psi_I)^2=1$. Commutation of two operators gives rise to a sign factor
\[
 \psi_{I_1}\psi_{I_2}=(-1)^k\psi_{I_2}\psi_{I_1},
\]
where $k=|I_1\cap I_2|$ is the number of indices contained in both $I_1$ and $I_2$. In the limit (\ref{ds}) the probability that $I_1$ and $I_2$ have $k$ indices in common is given by
\begin{equation}
p_k=\frac{1}{k!}\left(\frac\lambda2\right)^k\rme^{-\frac\lambda2}.
\end{equation}
Hence each time one commutes two $p$-fermion operators one receives a factor of
\begin{equation}
 \sum_{k=0}^\infty (-1)^kp_k=\rme^{-\lambda}\equiv q.
\end{equation}
To count how many times one needs to commute the $p$-fermion operators, it is convenient to think of a chord diagram in which the operators $\psi_{I_1},\cdots,\psi_{I_n}$ are put along the boundary of a disk in this order and each pair of operators is connected by a chord. A sample diagram is shown on the left of Figure \ref{fig:chord}. Then each time one commutes a pair of operators, the number of intersections of chords decreases by one. Thus
\begin{equation}
 m_n\;=\;\sum_{\text{diagrams}}q^{\#(\text{intersections})},
\end{equation}
where the sum is over different chord diagrams that specify the pairings of the $n$ points on the boundary of a disk.

\paragraph{Chord quantum mechanics}

The sum over chord diagrams can be reformulated as an amplitude of an auxiliary quantum mechanics which we call the {\em chord QM}. The idea is illustrated in the diagram on the right of Figure \ref{fig:chord} which is conformally equivalent to the one on the left. As is shown there, at each time step along the boundary of the upper half-plane, a chord is either emitted to or absorbed from the interior, and the number of chords changes by $\pm1$. So the Hamiltonian ${\sf H}$ of the chord QM is given by a sum of chord creation and annihilation operators ${\sf a}^{\!\dagger}, {\sf a}$. Then $Z(\beta)$ and $m_n$ can be expressed as
\begin{equation}
Z(\beta)=\langle0|\rme^{-\beta{\sf H}}|0\rangle,\qquad
m_n=\langle0|{\sf H}^n|0\rangle,\qquad
{\sf H}={\sf a}^{\!\dagger}+{\sf a},
\label{Zbmk}
\end{equation}
where $|0\rangle$ and $\langle0|$ are the states with no chords. General number eigenstates $|n\rangle$ and $\langle n|$ are defined to satisfy
\begin{equation}
\begin{aligned}
{\sf a}^{\!\dagger}|n\rangle &=|n+1\rangle,\\
{\sf a}|n\rangle  &=|n-1\rangle\,[n]_q,
\end{aligned}\qquad
\begin{aligned}
 \langle n|{\sf a}^{\!\dagger} &= \langle n-1|\,[n]_q,\\
 \langle n|{\sf a}\,\, &= \langle n+1|,
\end{aligned}\qquad
\langle n|n'\rangle = \delta_{nn'}\,[n]_q!\;.\quad
\left([n]_q!\equiv\sum_{k=1}^n[k]_q\right)
\end{equation}
The $q$-integer $[n]_q=1+q+\cdots+q^{n-1}$ appears here because, when one of $n$ chords is absorbed by the boundary, it has to cross some of the other $n-1$ chords. It follows from these rules that ${\sf a},{\sf a}^{\!\dagger}$ obey the Arik-Coon $q$-oscillator algebra \cite{Arik:1973vg}:
\begin{equation}
 {\sf a}{\sf a}^{\!\dagger} - q{\sf a}^{\!\dagger}{\sf a}\;=\; 1.
\label{ArikCoon}
\end{equation}
One also has ${\sf a}{\sf a}^{\!\dagger}-{\sf a}^{\!\dagger}{\sf a}=q^{\sf N}$, where ${\sf N}$ is the chord number operator.

Another state of importance is the eigenstate of ${\sf H}$ \cite{Berkooz:2018qkz}:
\begin{equation}
 {\sf H}|\theta\rangle = E(\theta)|\theta\rangle,\qquad
 E(\theta)=\frac{2\cos\theta}{\sqrt{1-q}}.
\end{equation}
As was pointed out in \cite{Berkooz:2018qkz}, $\langle n|\theta\rangle$ satisfies a recursion relation that is solved in terms of $q$-Hermite polynomials. One can also show that, if $|\theta\rangle$ is normalized so that $\langle 0|\theta\rangle=1$, it satisfies \cite{Carlitz:1957aa} (see Exercise 9.10 of \cite{Gasper:2004aa})
\begin{equation}
\langle\theta|t^{\sf N}|\theta'\rangle\;=\;
 \sum_{n=0}^\infty\frac{t^n}{[n]_q!}\langle\theta|n\rangle\langle n|\theta'\rangle\;=\;\frac{(t^2;q)_\infty}{(t\rme^{\rmi(\theta+\theta')},t\rme^{\rmi(\theta-\theta')},t\rme^{\rmi(-\theta+\theta')},t\rme^{\rmi(-\theta-\theta')};q)_\infty}.
\label{nrmth}
\end{equation}
Here we used the standard notation for $q$-Pochhammer symbol
\begin{equation}
 (a;q)_n=\prod_{j=0}^{n-1}(1-aq^j),\qquad
 (a_1,\cdots,a_m;q)_n=(a_1;q)_n\cdots(a_m;q)_n.
\end{equation}
By taking the limit $t\nearrow1$ of (\ref{nrmth}) one obtains (for $0\le\theta,\theta'\le\pi$)
\begin{equation}
\langle\theta|\theta'\rangle\;=\;
\frac{2\pi\delta(\theta-\theta')}{\mu(\theta)},\qquad\therefore\quad
1\;=\;\int_0^\pi\frac{\mathrm d\theta\mu(\theta)}{2\pi}|\theta\rangle\langle\theta|
\label{thbasis}
\end{equation}
with $\mu(\theta)\equiv(q,\rme^{2\rmi\theta},\rme^{-2\rmi\theta};q)_\infty$. Using $\langle0|\theta\rangle=1$ one can rewrite $Z(\beta)$ (\ref{Zbmk}) as follows:
\begin{equation}
 Z(\beta)\;=\;\int_0^\pi\frac{\mathrm d\theta \mu(\theta)\rme^{-\beta E(\theta)}}{2\pi}\;.
\end{equation}

\paragraph{Adding matter}

As a basic matter observable in the SYK model, we consider a polynomial of fermions of definite degree $\tilde p=p\Delta$:
\begin{equation}
 M_\mini{\Delta}\;\equiv\;\rmi^{\frac{\tilde p}2}\sum_{1\le i_1<\cdots<i_{\tilde p}\le N}\tilde J_{i_1\cdots i_{\tilde p}}\psi_{i_1}\cdots\psi_{i_{\tilde p}}.
\end{equation}
We take the disorder average with respect to the couplings $J_{i_1\cdots i_p}$ as well as $\tilde J_{i_1\cdots i_{\tilde p}}$ in the above, assuming that $\tilde J_{i_1\cdots i_{\tilde p}}$ obey Gaussian distribution with zero mean and
\begin{equation}
\big\langle\tilde J_{i_1\cdots i_{\tilde p}}\tilde J_{j_1\cdots j_{\tilde p}}\big\rangle_J\;=\;
\Big({}^N_{\,\tilde p}\Big)^{-1}\cdot\delta_{i_1j_1}\cdots\delta_{i_{\tilde p}j_{\tilde p}}.
\end{equation}

In the double-scaling limit
\begin{equation}
 N,p,\tilde p\to\infty\quad\text{with}\quad q\equiv \rme^{-\frac{2p^2}N}~~\text{and}~~\tilde q\equiv \rme^{-\frac{2p\tilde p}N}=q^\Delta\quad\text{fixed},
\end{equation}
the correlation functions of $M_\mini{\Delta}$ can be calculated by summing over diagrams made of two kinds of chords, namely the ``$H$-chords'' connecting two $H$'s and the ``$M$-chords'' connecting two $M_\mini{\Delta}$'s. Each intersection of two $H$-chords is weighted by a factor of $q$, while each intersection of an $H$-chord with an $M$-chord is weighted by $\tilde q$. The computation can again be reformulated in terms of the chord QM. For example, the two-point function of $M_\mini{\Delta}$ can be expressed as follows:
\begin{equation}
 G(\Delta|\beta_1,\beta_2)\;=\;\left\langle\mathrm{Tr}\big(\rme^{-\beta_1 H}M_\mini{\Delta}\rme^{-\beta_2H}M_\mini{\Delta}\big)\right\rangle_J\;=\;\langle 0|
\contraction{\rme^{-\beta_1{\sf H}}}{{\sf M}}{{}_\mini{\Delta}\rme^{-\beta_2{\sf H}}}{{\sf M}}
\rme^{-\beta_1{\sf H}}{\sf M}_\mini{\Delta}\rme^{-\beta_2{\sf H}}{\sf M}_\mini{\Delta}|0\rangle\,.
\label{2ptfn}
\end{equation}
Here and in what follows, in chord QM expressions we represent each $M$-chord by a contraction symbol and its endpoints by two ${\sf M}_\mini{\Delta}$'s. Note also that, while the thermal correlator in the original SYK model is defined by periodic identification of time, we need to cut open the time circle somewhere to rewrite it as a chord QM expression. It is a special property of the chord QM that the resulting amplitude does not depend on where the circle has been cut.

The following relation was shown in \cite{Berkooz:2018jqr} using diagrammatic argument:
\begin{equation}
\contraction{}{{\sf M}}{{}_\mini{\Delta}(\cdots)}{{\sf M}}
{\sf M}_\mini{\Delta}(\cdots){\sf M}_\mini{\Delta}\;=\;\sum_{n=0}^\infty\frac{(q^{2\Delta};q)_n}{[n]_q!}{\sf a}^{\!\dagger n}q^{\Delta \sf N}(\cdots)q^{\Delta\sf N}{\sf a}^n,
\label{defcnt}
\end{equation}
where $(\cdots)$ stands for any operator. An important consequence of this relation is that the operator $\contraction{}{{\sf M}}{{}_\mini{\Delta}(\cdots)}{{\sf M}}{\sf M}_\mini{\Delta}(\cdots){\sf M}_\mini{\Delta}$ commutes with ${\sf H}$ as long as the operator $(\cdots)$ does. This can be proven easily by using (\ref{defcnt}) together with
\begin{align}
{\sf H}{\sf a}^{\!\dagger\,n}q^{\Delta{\sf N}}&\;=\;
q^{n+\Delta}{\sf a}^{\!\dagger\,n}q^{\Delta{\sf N}}{\sf H}
+(1-q^{n+2\Delta}){\sf a}^{\!\dagger\, n+1}q^{\Delta{\sf N}}
+[n]_q{\sf a}^{\!\dagger\,n-1}q^{\Delta{\sf N}}, \nonumber \\
q^{\Delta{\sf N}}{\sf a}^n{\sf H}&\;=\;
q^{n+\Delta}{\sf H}\,q^{\Delta{\sf N}}{\sf a}^n\;
+(1-q^{n+2\Delta})q^{\Delta{\sf N}}{\sf a}^{n+1}\;
+[n]_qq^{\Delta{\sf N}}{\sf a}^{n-1}.
\end{align}
This fact is useful in making sure that the chord QM amplitudes do not depend on where to cut open the time circle. By substituting (\ref{defcnt}) into (\ref{2ptfn}), using the completeness of the bases $|n\rangle,\;|\theta\rangle$ and recalling (\ref{nrmth}) one obtains an exact formula for the two-point function:
\begin{align}
G(\Delta|\beta_1,\beta_2)&\;=\;\int_0^\pi\prod_{i=1,2}\frac{\mathrm d\theta_i\mu(\theta_i)\rme^{-\beta_iE(\theta_i)}}{2\pi}G(\Delta|\theta_1,\theta_2), \nonumber\\
G(\Delta|\theta_1,\theta_2)&\;=\;
\langle\theta_1|
\contraction{}{{\sf M}}{{}_\mini{\Delta}|\theta_2\rangle\langle\theta_2|}{{\sf M}}
{\sf M}_\mini{\Delta}|\theta_2\rangle\langle\theta_2|{\sf M}_\mini{\Delta}
|0\rangle \;=\;\langle\theta_1|q^{\Delta{\sf N}}|\theta_2\rangle \nonumber \\[1mm] &
\;=\; \frac{(q^{2\Delta};q)_\infty}{(q^\Delta \rme^{\rmi(\theta_1+\theta_2)},q^\Delta \rme^{\rmi(\theta_1-\theta_2)},q^\Delta \rme^{\rmi(-\theta_1+\theta_2)},q^\Delta \rme^{\rmi(-\theta_1-\theta_2)};q)_\infty}\,.
\label{GDelta}
\end{align}
Before proceeding, let us note that, using $\langle0|\theta\rangle=1$ as well as the fact that $|\theta\rangle\langle\theta|$ commutes with any operator that commutes with ${\sf H}$, one can derive the following useful relation:
\begin{equation}
\langle\theta_0|
\contraction{}{{\sf M}}{{}_\mini{\Delta}|\theta_1\rangle\langle\theta_1|}{{\sf M}}{\sf M}_\mini{\Delta}|\theta_1\rangle\langle\theta_1|{\sf M}_\mini{\Delta}|\theta_2\rangle
\;=\;
\langle\theta_0|\theta_2\rangle\langle\theta_2|
\contraction{}{{\sf M}}{{}_\mini{\Delta}|\theta_1\rangle\langle\theta_1|}{{\sf M}}{\sf M}_\mini{\Delta}|\theta_1\rangle\langle\theta_1|{\sf M}_\mini{\Delta}|0\rangle
\;=\;
\frac{2\pi\delta(\theta_0-\theta_2)}{\mu(\theta_0)}G(\Delta|\theta_1,\theta_2).
\label{bldiag}
\end{equation}

\begin{figure}[t]
\begin{center}
\begin{overpic}[scale=1]{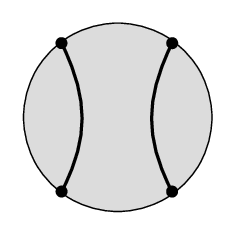}
\put(18,87){\footnotesize $\sf M_\mini{\Delta}$}
\put(18,10){\footnotesize $\sf M_\mini{\Delta}$}
\put(75,10){\footnotesize $\sf M_\mini{\Delta'}$}
\put(73,86){\footnotesize $\sf M_\mini{\Delta'}$}
\put(47,93){\footnotesize $\beta_0$}
\put( 1,47){\footnotesize $\beta_1$}
\put(47, 2){\footnotesize $\beta_2$}
\put(92,47){\footnotesize $\beta_3$}
\put(18,47){\footnotesize $\theta_1$}
\put(47,47){\footnotesize $\theta_2$}
\put(76,47){\footnotesize $\theta_3$}
\end{overpic}
\end{center}
\caption{An uncrossed four-point function (\ref{4ptfn2}) and the parameters $\beta_i,\theta_i$.}\label{fig:unc4}
\end{figure}
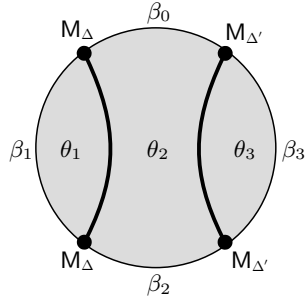

In \cite{Berkooz:2018jqr}, the diagrammatic analysis was extended to some higher-point correlation functions. In particular, an exact formula for the {\it crossed} four-point functions, i.e. correlators with crossings of matter chords, was obtained, and its relation to the $6j$ symbol of the quantum group ${\cal U}_{q^{1/2}}({\mathfrak{su}}_{1,1})$ was pointed out. The results obtained there were summarized into a set of diagrammatic rules for general correlators. In this paper we will need a formula for the {\it uncrossed} four-point functions:
\begin{align}
G(\Delta,\Delta'|\beta_0,\beta_1,\beta_2,\beta_3)&\;=\;
\left\langle\mathrm{Tr}\big(
\rme^{-\beta_0 H}M_\mini{\Delta}\rme^{-\beta_1H}M_\mini{\Delta}
\rme^{-\beta_2 H}M_\mini{\Delta'}\rme^{-\beta_3H}M_\mini{\Delta'}
\big)\right\rangle_J \nonumber \\ &\;=\;\langle 0|
\contraction{\rme^{-\beta_0{\sf H}}}{{\sf M}}{{}_\mini{\Delta}\rme^{-\beta_1{\sf H}}}{{\sf M}}
\rme^{-\beta_0{\sf H}}{\sf M}_\mini{\Delta}\rme^{-\beta_1{\sf H}}{\sf M}_\mini{\Delta}
\contraction{\rme^{-\beta_2{\sf H}}}{{\sf M}}{{}_\mini{\Delta'}\rme^{-\beta_3{\sf H}}}{{\sf M}}
\rme^{-\beta_2{\sf H}}{\sf M}_\mini{\Delta'}\rme^{-\beta_3{\sf H}}{\sf M}_\mini{\Delta'}
|0\rangle\,.
\label{4ptfn}
\end{align}
By inserting complete sets of ${\sf H}$-eigenstates and then using (\ref{bldiag}), one can rewrite it in terms of the two-point functions (\ref{GDelta}):
\begin{align}
&G(\Delta,\Delta'|\beta_0,\beta_1,\beta_2,\beta_3)\;=\;
\int_0^\pi\prod_{i=0}^3\frac{\mathrm d\theta_i\mu(\theta_i)\rme^{-\beta_iE(\theta_i)}}{2\pi}
\langle\theta_0|
\contraction{}{{\sf M}}{{}_\mini{\Delta}|\theta_1\rangle\langle\theta_1|}{{\sf M}}
{\sf M}_\mini{\Delta}|\theta_1\rangle\langle\theta_1|{\sf M}_\mini{\Delta}|\theta_2\rangle
\contraction{\langle\theta_2|}{{\sf M}}{{}_\mini{\Delta'}|\theta_3\rangle\langle\theta_3|}{{\sf M}'}
\langle\theta_2|{\sf M}_\mini{\Delta'}|\theta_3\rangle\langle\theta_3|{\sf M}_\mini{\Delta'}|0\rangle
\nonumber\\
&~\hskip20mm~\;=\;
\int_0^\pi\prod_{i=1}^3\frac{\mathrm d\theta_i\mu(\theta_i)}{2\pi}
\rme^{-\beta_1E(\theta_1)-(\beta_0+\beta_2)E(\theta_2)-\beta_3E(\theta_3)}
G(\Delta|\theta_1,\theta_2)G(\Delta'|\theta_2,\theta_3).
\label{4ptfn2}
\end{align}
Figure \ref{fig:unc4} represents the four-point function considered here. Note that, while $\beta_0,\cdots,\beta_3$ are the lengths of the boundary arcs, the parameters $\theta_1,\theta_2,\theta_3$ can be thought of as assigned to the three regions of the disk divided by matter chords.

\section{Boundary Liouville CFT}\label{sec:BLCFT}

Here we make a slight detour and give a brief introduction to boundary Liouville CFT. This is because our analysis of the DSSYK model in Section \ref{sec:deg} borrows many ideas from the work \cite{Fateev:2000ik} on boundary Liouville correlation functions, and the results have some similarity as well.

Liouville CFT is a 2d theory of a single scalar field $\phi$. The action on a surface $\Sigma$ with boundary $\partial\Sigma$ is given by
\begin{equation}
 S\;=\;\int_\Sigma\frac{\mathrm d^2x\sqrt{g}}{4\pi}\left\{g^{mn}\partial_m\phi\partial_n\phi+QR\phi+4\pi\mu\rme^{2b\phi}\right\}+\int_{\partial\Sigma}\frac{\mathrm dx g^{1/4}}{2\pi}\left\{QK\phi+2\pi\mu_\text{B}\rme^{b\phi}\right\},
\label{Sliou}
\end{equation}
where $g_{mn}$ is the metric on $\Sigma$ and $g$ is its determinant. $R$ is the scalar curvature of $\Sigma$ and $K$ is the extrinsic curvature of $\partial\Sigma$, in terms of which the Euler characteristic of a surface with $g$ handles and $h$ holes is given by
\begin{equation}
\int_\Sigma\frac{\mathrm d^2x\sqrt{g}R}{4\pi}+\int_{\partial\Sigma}\frac{\mathrm dx g^{1/4}K}{2\pi}\;=\; 2-2g-h.
\end{equation}
The parameter $b$ is called the Liouville coupling, and $Q=b+\frac1b$. The theory is a CFT with central charge $c=1+6Q^2$, and it is known to be self-dual under $b\leftrightarrow\frac1b$.

The cosmological constant $\mu$ sets the scale of the theory, while the boundary cosmological constant $\mu_\text{B}$ can take different values according to the choice of conformally invariant boundary conditions (D-branes) for each boundary segment. $\mu_\text{B}$ is related to the label $s$ of the so-called FZZT-branes by
\begin{equation}
 \mu_{\text{B}[s]}\;=\;\sqrt{\frac{\mu}{\sin\pi b^2}}\cosh 2\pi bs\,.
\label{muB}
\end{equation}
Note that the FZZT-branes with labels $s$ and $(-s)$ are equivalent.

The operator $V_\alpha=\rme^{2\alpha\phi}$ inserted in the bulk of $\Sigma$ is a Virasoro primary of conformal weight $\alpha(Q-\alpha)$, whereas the operator $B_\beta=\rme^{\beta\phi}$ inserted on $\partial\Sigma$ is a boundary Virasoro primary of weight $\beta(Q-\beta)$. The goal of this section is to explain the derivation of the two-point function of boundary operators. Let us put two $B_\beta$'s on the boundary of the upper half-plane ($x$-axis), and label the FZZT-branes for the two boundary components by $s_1,s_2$. Then \cite{Fateev:2000ik}
\begin{align}
\langle B_{\beta_1}(x_1)B_{\beta_2}(s_2)\rangle_{s_1,s_2}&\;=\;
|x_1-x_2|^{-2\beta_1(Q-\beta_1)}\Big\{
\delta\big(\rmi(\beta_1+\beta_2-Q)\big)+\delta\big(\rmi(\beta_1-\beta_2)\big)d(\beta|s_1,s_2)\Big\}, \nonumber \\
d(\beta|s_1,s_2)&\;=\;\frac
{(\pi\mu\gamma(b^2)b^{2-2b^2})^{\frac{Q-2\beta}{2b}}\bG(Q-2\beta)\bG(2\beta-Q)^{-1}}
{\bS(\beta+\rmi s_1+\rmi s_2)\bS(\beta+\rmi s_1-\rmi s_2)\bS(\beta-\rmi s_1+\rmi s_2)\bS(\beta-\rmi s_1-\rmi s_2)}\,,
\label{LB2pt}
\end{align}
where $\gamma(x)=\Gamma(x)/\Gamma(1-x)$ and we used the functions $\bG(x)$ and $\bS(x)=\bG(Q-x)/\bG(x)$ introduced in \cite{Fateev:2000ik}. They satisfy the shift relations
\begin{equation}
\begin{alignedat}{2}
\bG(x+b)&\;=\; \frac{b^{\frac12-bx}}{\sqrt{2\pi}}\Gamma(bx)\,\bG(x),\qquad&
\bG(x+\tfrac1b)&\;=\; \frac{b^{\frac xb-\frac12}}{\sqrt{2\pi}}\Gamma(\tfrac xb)\,\bG(x),\\
\bS(x+b)&\;=\; 2\sin\pi bx\,\bS(x),\qquad&
\bS(x+\tfrac1b)&\;=\; 2\sin\tfrac{\pi x}b\,\bS(x),
\end{alignedat}
\label{bGbS}
\end{equation}
and $\bG(x)$ has zeroes at $x=-mb-nb^{-1}~(m,n\in\mathbb{Z}_{\ge0})$. The two delta functions in (\ref{LB2pt}) ensure that $B_{\beta_1},B_{\beta_2}$ have equal conformal weights when $\beta_1,\beta_2\in\frac Q2+\rmi\mathbb R$. Note that (\ref{LB2pt}) implies that the boundary operators between two FZZT-branes $s_1,s_2$ obey an equivalence relation
\begin{equation}
 [B_\beta]_{s_1,s_2} = d(\beta|s_1,s_2)[B_{Q-\beta}]_{s_1,s_2}\,,
\end{equation}
from which it also follows that $d(\beta|s_1,s_2)d(Q-\beta|s_1,s_2)=1$.

\paragraph{Properties of degenerate operators}

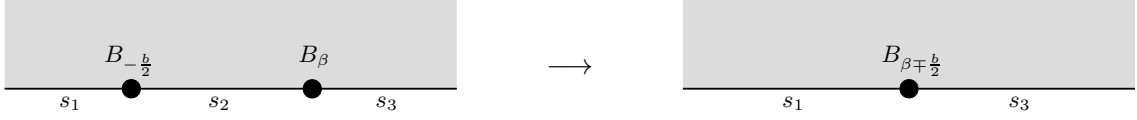
\begin{figure}[t]
\begin{center}
\begin{tabular}{ccc}
\begin{overpic}[scale=1.2]{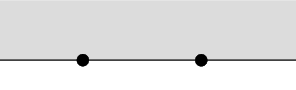}
\put(22,16){\footnotesize $B_{-\frac b2}$}
\put(65,16){\footnotesize $B_\beta$}
\put(12, 6){\footnotesize $s_1$}
\put(45, 6){\footnotesize $s_2$}
\put(82, 6){\footnotesize $s_3$}
\end{overpic}
&\raisebox{8mm}{$\qquad\longrightarrow\qquad$}&
\begin{overpic}[scale=1.2]{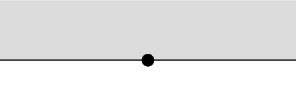}
\put(44,16){\footnotesize $B_{\beta\mp\frac b2}$}
\put(22, 6){\footnotesize $s_1$}
\put(72, 6){\footnotesize $s_3$}
\end{overpic}
\end{tabular}
\end{center}
\caption{OPE of boundary operators.}\label{fig:bOPE}
\end{figure}

The derivation of $d(\beta|s_1,s_2)$ in \cite{Fateev:2000ik} uses the boundary OPE relations which involve special operators corresponding to degenerate representations of Virasoro algebra, i.e., representations with null states. In this paper we use the most basic such operator $B_{-\frac b2}$ that satisfies
\begin{equation}
 \Big(\partial^2+b^2T(x)\Big)B_{-\frac b2}(x)=0,
\label{nullv}
\end{equation}
where $T=-(\partial\phi)^2+Q\partial^2\phi$ is the stress tensor. The corresponding Virasoro representation has a null state
\begin{equation}
(L_{-1}^2+b^2L_{-2})|{\beta=-\tfrac b2}\rangle
\label{nullv2}
\end{equation}
at level two. Using this one can show that the product $B_{-\frac b2}B_\beta$ can be expanded into a linear combination of two primaries $B_{\beta-\frac b2}, B_{\beta+\frac b2}$ and their descendants as follows:
\begin{align}
 [B_{-\frac b2}(x_1)]_{s_1,s_2}[B_\beta(x_2)]_{s_2,s_3}\;\xrightarrow{~x_1\to x_2~}~~\;&
 c_+\cdot|x_1-x_2|^{b\beta}[B_{\beta-\frac b2}(x_2)]_{s_1,s_3}\nonumber\\
 +\;& c_-\cdot|x_1-x_2|^{b(Q-\beta)}[B_{\beta+\frac b2}(x_2)]_{s_1,s_3}.
\label{bOPE}
\end{align}
This OPE relation is illustrated in Figure \ref{fig:bOPE}.

Another important property of $B_{-\frac b2}$ is that it exists between two FZZT branes $s_1$ and $s_2$ only when $s_1-s_2=\pm\frac{\rmi b}2$ or $s_1+s_2=\pm\frac{\rmi b}2$. This is because the FZZT-brane $s$ is related to the Virasoro representation of conformal weight $s^2+\frac{Q^2}4$ (and momentum $\frac Q2+\rmi s$) via modular bootstrap \cite{Zamolodchikov:2001ah}. The spectrum of boundary operators between D-branes must be consistent with the fusion rule \cite{Cardy:1986gw,Cardy:1989ir}, namely $B_\beta$ can exist between the FZZT-branes $s_1,s_2$ only when the fusion product of two representations $[\frac Q2+\rmi s_1]\times[\beta]$ contains $[\frac Q2+\rmi s_2]$.\label{condss}

\paragraph{Path integral argument}

The coefficients $c_\pm$ in (\ref{bOPE}) can be computed by making use of the following fact. For a general correlator of operators $V_{\alpha_i}$ and $B_{\beta_j}$ on a surface $\Sigma$ with $g$ handles and $h$ boundaries, let us define the total Liouville momentum $P$ conjugate to the shift $\phi\to\phi+\text{constant}$ as follows:
\begin{equation}
 P ~\equiv~ \sum_i2\alpha_i+\sum_j\beta_j-Q(2-2g-h).
\end{equation}
Note that the last term in the RHS is the momentum of the background curvature which arises from the linear dilaton coupling in (\ref{Sliou}). Since the shift symmetry is explicitly broken by cosmological terms, correlators are non-vanishing even when $P\ne0$. However, a useful fact is that the correlator has simple poles at $P=-nb-mb^{-1}$ with $m,n$ non-negative integers. Moreover, the residue at these poles can be evaluated as a Wick contraction of free theory with the insertion of a suitable number of cosmological terms to screen $P$ \cite{Feigin:1981st,Dotsenko:1984ad}. To understand how this works in a simple example, consider the path integral for a correlator $\langle\prod_iV_{\alpha_i}(z_i)\rangle$ on a closed surface of genus $g$. By integrating over the constant mode $\phi_0$ of the Liouville field first, one would obtain \cite{Teschner:2001rv}
\begin{equation}
\int\mathrm d\phi_0\exp\left(P\phi_0-\rme^{2b\phi_0}\mu\int\widehat V_b\right)
\;=\;\frac{\Gamma(\frac P{2b})}{2b}\Big(\mu{\textstyle\int\widehat V_b}\Big)^{-\frac P{2b}}
\;\simeq\;\sum_{n\ge0}\frac{(-\mu\int\widehat V_b)^n/n!}{P+2nb},
\end{equation}
where $P=2\sum_i\alpha_i-Q(2-2g)$ and the hat on $V_b$ indicates that its $\phi_0$-dependence has been removed. This indeed captures part of the poles explained above.  Then, to evaluate the residue at the pole $P=-2nb$, one needs to perform the path integral
\begin{equation}
\int{\cal D}\widehat\phi\,\rme^{-\widehat S_\text{LD}}\,\frac1{n!}\Big({-\mu}{\textstyle\int \widehat V_b}\Big)^n\prod_{i}\widehat V_{\alpha_i}(z_i)\;=:\;
\bigg\langle\frac1{n!}\Big({-\mu}{\textstyle\int \widehat V_b}\Big)^n\prod_{i}\widehat V_{\alpha_i}(z_i)\bigg\rangle_\text{Wick}
\end{equation}
over the remaining non-constant modes with the action $\widehat S_\text{LD}$ for a free linear-dilaton CFT. This is indeed given by the Wick contraction with $n$ additional insertion of cosmological operators.

Now, by multiplying $B_{Q-\beta+\frac b2}(x_3)$ to both sides of (\ref{bOPE}) and evaluating correlators, one finds that $c_+$ is given by the ratio:
\begin{align}
c_+(\beta|s_1,s_2,s_3) &\;=\; \lim_{x_3\to\infty}
\frac{\big\langle B_{-\frac b2}(x_1)B_\beta(x_2)B_{Q-\beta+\frac b2}(x_3)\big\rangle}
{|x_1-x_2|^{b\beta}\cdot\big\langle B_{\beta-\frac b2}(x_2)B_{Q-\beta+\frac b2}(x_3)\big\rangle}.
\end{align}
The correlators in the denominator and the numerator are both divergent (as they both have $P=0$), so one may replace them by the residues at the pole $P=0$:
\begin{align}
c_+(\beta|s_1,s_2,s_3) &\;=\; \lim_{x_3\to\infty}
\frac{\big\langle \widehat B_{-\frac b2}(x_1)\widehat B_\beta(x_2)\widehat B_{Q-\beta+\frac b2}(x_3)\big\rangle_\text{Wick}}
{|x_1-x_2|^{b\beta}\cdot\big\langle\widehat B_{\beta-\frac b2}(x_2)\widehat B_{Q-\beta+\frac b2}(x_3)\big\rangle_\text{Wick}} \nonumber \\ &\;=\;\lim_{x_3\to\infty}
\frac{|x_1-x_2|^{b\beta}|x_1-x_3|^{b(Q-\beta+\frac b2)}|x_2-x_3|^{-2\beta(Q-\beta+\frac b2)}}{|x_1-x_2|^{b\beta}\cdot|x_2-x_3|^{(\beta-\frac b2)(Q-\beta+\frac b2)}}\;=\;1.
\end{align}
Here the Wick contraction was evaluated using $\phi(x_1)\phi(x_2)\sim -2\ln|x_1-x_2|$. Similarly, one can express $c_-$ as a ratio of divergent correlators, but then the correlator in the numerator has $P=-1$. Therefore, it is replaced by the Wick contraction with one boundary cosmological term inserted.
\begin{align}
c_-(\beta|s_1,s_2,s_3) &\;=\; \lim_{x_3\to\infty}
\frac{\displaystyle\int\mathrm dx\big\langle\big({-\mu_\text{B}(x)}\widehat B_b(x)\big)\widehat B_{-\frac b2}(x_1)\widehat B_\beta(x_2)\widehat B_{Q-\beta-\frac b2}(x_3)\big\rangle_\text{Wick}}
{|x_1-x_2|^{b(Q-\beta)}\cdot\big\langle \widehat B_{\beta+\frac b2}(x_2)\widehat B_{Q-\beta-\frac b2}(x_3)\big\rangle_\text{Wick}}.
\label{c-1}
\end{align}
Here $\mu_\text{B}(x)$ equals $\mu_{\text{B}[s_1]},\mu_{\text{B}[s_2]}$ or $\mu_{\text{B}[s_3]}$ depending on the position of $\widehat B_b(x)$. By setting $(x_1,x_2,x_3)$ to $(0,1,\infty)$ and performing the Wick contraction one obtains
\begin{align}
c_-(\beta|s_1,s_2,s_3) &\;=\;
-\int\mathrm dx\mu_\text{B}(x)|x|^{b^2}|1-x|^{-2b\beta}
\nonumber\\[1mm] &\;=\;
\frac{\Gamma(1+b^2)\Gamma(1-2b\beta)\Gamma(2b\beta-b^2-1)}\pi\nonumber \\ &\hskip12mm\cdot\Big\{
{-\mu}_{\text{B}[s_1]}\sin2\pi b\beta
+\mu_{\text{B}[s_2]}\sin\pi(2b\beta-b^2)
+\mu_{\text{B}[s_3]}\sin\pi b^2
\Big\}\,.
\label{c-2}
\end{align}
For $s_1=s_2\mp\frac{\rmi b}2$ the above expression can be factorized into the following form:
\begin{align}
 c_-(\beta|s_1,s_2,s_3)\Big|_{s_1=s_2\mp\rmi b/2}\;=\;&\frac{b^2\sqrt{\mu\pi\gamma(b^2)}}{2\pi}\Gamma(2b\beta-b^2-1)\Gamma(1-2b\beta)\nonumber \\ &\cdot 2\sin\pi b(\beta\pm\rmi s_2+\rmi s_3)\cdot 2\sin\pi b(\beta\pm\rmi s_2-\rmi s_3).
\label{c-3}
\end{align}

\paragraph{Recursion relation for $d(\beta|s_1,s_2)$}

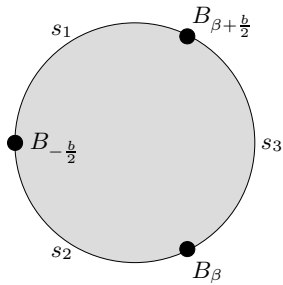
\begin{figure}[t]
\begin{center}
\begin{overpic}[scale=1]{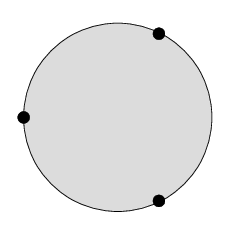}
\put(69,90){\footnotesize $B_{\beta+\frac b2}$}
\put(69, 5){\footnotesize $B_{\beta}$}
\put(15,48){\footnotesize $B_{-\frac b2}$}
\put(22,86){\footnotesize $s_1$}
\put(22,12){\footnotesize $s_2$}
\put(92,48){\footnotesize $s_3$}
\end{overpic}
\end{center}
\caption{A disk correlator that can be expressed in terms of $c_\pm$ and $d$ coefficients.}\label{fig:threepts}
\end{figure}

Finally, consider the disk three-point function illustrated in Figure \ref{fig:threepts}. Its dependence on the position of the three operators is determined completely from the conformal invariance. The coefficient can be expressed in terms of $c_\pm$ and $d$, but one finds two different ways to do so by sending $B_{-\frac b2}$ towards $B_{\beta+\frac b2}$ or $B_\beta$. This leads to a nontrivial recursion relation for $d(\beta|s_1,s_2)$ (\ref{LB2pt}):
\begin{equation}
c_+(\beta+\tfrac b2|s_2,s_1,s_3)d(\beta|s_2,s_3)\;=\; c_-(\beta|s_1,s_2,s_3)d(\beta+\tfrac b2|s_1,s_3).
\qquad\Big(s_1=s_2\mp\tfrac{\rmi b}2\Big)
\end{equation}
A similar recursion relation which shifts the parameters $\beta$ and $\rmi s$ by $\pm\frac1{2b}$ can be derived by using the other basic degenerate operator $B_{-\frac1{2b}}$. The formula (\ref{LB2pt}) for $d(\beta|s_1,s_2)$ can be easily reproduced by solving these recursion relations with the condition $d(\frac Q2|s_1,s_2)=1$.

\section{Degenerate matter operators in DSSYK}\label{sec:deg}

The DSSYK model and the Liouville CFT are both known to have some relation to ${\cal U}_{q^{1/2}}(\mathfrak{su}_{1,1})$ quantum group. In Liouville CFT, an example where this relation shows up is the equivalence between its fusion coefficient and the $6j$ symbol of the quantum group. This fact was applied to the verification of Liouville bootstrap \cite{Ponsot:1999uf,Ponsot:2000mt} and the determination of the boundary three-point function \cite{Ponsot:2001ng}. Note that the parameters $\beta,s$ are related to the labels of quantum group representations; in particular, the special properties of the operator $B_{-\frac b2}$ reflect that it corresponds to the doublet representation.

It would then be natural to ask whether the DSSYK model has similar degenerate operators with special properties. In this section we find them and clarify their basic properties by using only the facts reviewed in Section \ref{sec:DSSYK} and no other prerequisite knowledge.

\paragraph{Finding special values of $\Delta$}

We first ask if there are special values of $\Delta$ such that $\langle\theta|{\sf M}_\mini{\Delta}|\theta'\rangle$ is non-vanishing only when $\theta$ and $\theta'$ obey some conditions. Our analysis here is similar to that of \cite{Mertens:2020pfe} which studied the behavior of degenerate operators in JT and Liouville gravities through disk two-point functions.

We know that when $\Delta=0$ and therefore ${\sf M}_\mini{\Delta}=1$, one has $\langle\theta|{\sf M}_\mini{\Delta}|\theta'\rangle=\langle\theta|\theta'\rangle$ which vanishes unless $\theta=\theta'$. As was already mentioned in (\ref{thbasis}), this behavior can be reproduced from the general formula (\ref{GDelta}) for $G(\Delta|\theta,\theta')$. Indeed, the factor $(q^{2\Delta};q)_\infty$ in the numerator vanishes as $\Delta\to 0$, but the denominator also vanishes if $\theta$ happens to equal $\theta'$. In fact, $G(\Delta|\theta,\theta')$ behaves near $\theta=\theta'$ as
\begin{equation}
G(\Delta|\theta,\theta')~\stackrel{\Delta\searrow0}\longrightarrow~
\frac{(1-q^{2\Delta})}{(1-q^{\Delta}\rme^{\rmi(\theta-\theta')})(1-q^{\Delta}\rme^{-\rmi(\theta-\theta')})}\frac{(q;q)_\infty}{(q,q,\rme^{\rmi(\theta+\theta')},\rme^{-\rmi(\theta+\theta')};q)_\infty}\;=\;\frac{2\pi\delta(\theta-\theta')}{\mu(\theta')}\,.
\end{equation}
We see that the delta function arises as a result of a zero in the numerator and two colliding zeroes in the denominator.

The first nontrivial value of $\Delta$ for which $G(\Delta|\theta,\theta')$ becomes delta-functional is $\Delta=-\frac12$:
\begin{equation}
G(\Delta|\theta,\theta')~\stackrel{\Delta\searrow-\frac12}\longrightarrow~
\frac{2\pi\delta(\theta+\frac{\rmi\lambda}2-\theta')}{\mu(\theta')}\frac1{1-\rme^{2\rmi\theta}}+
\frac{2\pi\delta(\theta-\frac{\rmi\lambda}2-\theta')}{\mu(\theta')}\frac1{1-\rme^{-2\rmi\theta}}.
\label{G-1/2}
\end{equation}
Here we omitted the terms proportional to $\delta(\theta+\theta'\pm\frac{\rmi\lambda}2)$, but they can be easily worked out from the fact that $G(\Delta|\theta,\theta')$ is an even function in $\theta,\theta'$. This property of ${\sf M}_\mini{-\frac12}$ is similar to that the Liouville boundary operator $B_{-b/2}$ can only exist between the FZZT-branes $s_1,s_2$ obeying suitable condition. Note that such an ${\sf M}_\mini{-\frac12}$ can only be defined by analytic continuation in $\Delta$, as it would correspond to a product of negative number of Majorana fermions in the original SYK model.

\paragraph{Null vector equation}

Recall that the FZZT-branes $s_1,s_2$ on the two sides of $B_{-\frac b2}$ are related because $B_{-\frac b2}$ corresponds to a degenerate Virasoro representation with a null vector (\ref{nullv2}). We now claim that ${\sf M}_\mini{-\frac12}\equiv {\sf D}$ also satisfies a kind of null vector equation:
\begin{equation}
 \chi({\sf D})\;\equiv\; {\sf H}^2{\sf D}-(q^{\frac12}+q^{-\frac12}){\sf HDH}+{\sf D}{\sf H}^2+(q^{-1}-1){\sf D}\;=\;0.
\label{nullvec}
\end{equation}
This form can be found by requiring that its matrix element between ${\sf H}$-eigenstates $\langle\theta|$ and $|\theta'\rangle$ vanishes precisely when $\theta-\theta'=\pm\frac{\rmi\lambda}2$ or $\theta+\theta'=\pm\frac{\rmi\lambda}2$:
\begin{align}
\langle\theta|\chi({\sf D})|\theta'\rangle &\;=\;
\langle\theta|{\sf D}|\theta'\rangle\cdot\Big\{E(\theta)^2-(q^{\frac12}+q^{-\frac12})E(\theta)E(\theta')+E(\theta')^2+q^{-1}-1\Big\}
\nonumber \\ &\;=\;\frac{\langle\theta|{\sf D}|\theta'\rangle}{q(1-q)}\cdot
(\rme^{\rmi\theta+\rmi\theta'}-q^{\frac12})
(\rme^{\rmi\theta-\rmi\theta'}-q^{\frac12})
(\rme^{-\rmi\theta+\rmi\theta'}-q^{\frac12})
(\rme^{-\rmi\theta-\rmi\theta'}-q^{\frac12}).
\end{align}
So, $G(-\frac12|\theta,\theta')$ is delta-functional simply because $\chi({\sf D})$ is null.

Here we give a diagrammatic proof that $\chi({\sf D})$ is null. Let us think of chord diagrams in which one $\chi({\sf D})$ is inserted somewhere along the boundary of the upper half-plane. The operator $\chi({\sf D})$ may source an $M$-chord as well as two $H$-chords, or an $M$-chord only. We look closely into the vicinity of the $\chi({\sf D})$ insertion assuming that a $D$-chord extends from it upwards.

Consider first the situation where $\chi({\sf D})$ sources two $H$-chords in addition to the $D$-chord and they both extend to the left of the $D$-chord. We list up all the different chord diagrams in which the $2+1$ chords start from $\chi({\sf D})$ and intersect with each other before extending in the designated directions. By summing over all such diagrams one finds\vskip1mm
\begin{align}
\raisebox{-3mm}{\begin{overpic}[scale=1.4]{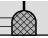}
\put(37,85){\scriptsize $\sf D$}
\put(-25,12){\scriptsize $\sf H$}
\put(-25,40){\scriptsize $\sf H$}
\put(23,-20){\scriptsize $\chi(\sf D)$}
\end{overpic}}
&\;=\;
\left(\;
 \raisebox{-3mm}{\includegraphics[scale=1.4]{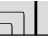}}
+\raisebox{-3mm}{\includegraphics[scale=1.4]{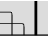}}
\;\right)
-(q^{\frac12}+q^{-\frac12})
\left(\;
 \raisebox{-3mm}{\includegraphics[scale=1.4]{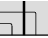}}
+\raisebox{-3mm}{\includegraphics[scale=1.4]{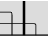}}
\;\right)
+\left(\;
 \raisebox{-3mm}{\includegraphics[scale=1.4]{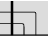}}
+\raisebox{-3mm}{\includegraphics[scale=1.4]{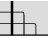}}
\;\right)
\nonumber \\[2mm] &\;=\; (1+q)-(q^{\frac12}+q^{-\frac12})(\tilde q+q\tilde q)
+(\tilde q^2+\tilde q^2q)\;=\; 0,\qquad(\tilde q=q^{-\frac12})
\end{align}
which indicates that $\chi({\sf D})$ is null. Let us examine all the other cases where the two $H$-chords extend in different directions. If there are two $H$-chords both extending rightwards, the sum over the diagrams gives the same result as above. If one $H$-chord extends to the left and another one to the right, the diagram sum becomes
\begin{align}
\raisebox{-3mm}{\includegraphics[scale=1.4]{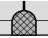}}
&\;=\;
\left(\;
 \raisebox{-3mm}{\includegraphics[scale=1.4]{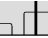}}
+\raisebox{-3mm}{\includegraphics[scale=1.4]{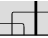}}
\;\right)
-(q^{\frac12}+q^{-\frac12})
\left(\;
 \raisebox{-3mm}{\includegraphics[scale=1.4]{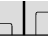}}
+\raisebox{-3mm}{\includegraphics[scale=1.4]{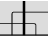}}
\;\right)
+\left(\;
 \raisebox{-3mm}{\includegraphics[scale=1.4]{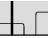}}
+\raisebox{-3mm}{\includegraphics[scale=1.4]{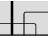}}
\;\right)
\nonumber \\[2mm] &\;=\; (\tilde q+q\tilde q)-(q^{\frac12}+q^{-\frac12})(1+q\tilde q^2)
+(\tilde q+\tilde qq)\;=\; 0.
\end{align}
Finally, the sum over diagrams with no outgoing $H$-chords is given by
\begin{align}
\raisebox{-3mm}{\includegraphics[scale=1.4]{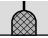}}
&\;=\;
 \raisebox{-3mm}{\includegraphics[scale=1.4]{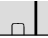}}
\;-\;(q^{\frac12}+q^{-\frac12})\cdot
 \raisebox{-3mm}{\includegraphics[scale=1.4]{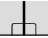}}
\;+\;\raisebox{-3mm}{\includegraphics[scale=1.4]{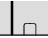}}
\;+\;(q^{-1}-1)\cdot
 \raisebox{-3mm}{\includegraphics[scale=1.4]{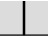}}
\nonumber \\[2mm] &\;=\; 1-(q^{\frac12}+q^{-\frac12})\cdot \tilde q
+1+(q^{-1}-1)\;=\; 0.
\end{align}
This completes the proof.

\paragraph{Two-sided formalism}

\begin{figure}[b]
\begin{center}
\begin{overpic}[scale=1.2]{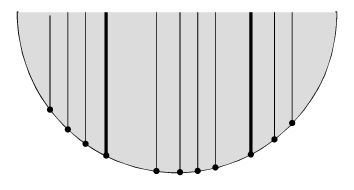}
\put(28, 5){\footnotesize $\Delta_1$}
\put(69, 5){\footnotesize $\Delta_2$}
\end{overpic}
\end{center}
\caption{A half-disk with two matter chords and $3+4+2$ $H$-chords, corresponding to a basis vector $|3,4,2\rangle$ in the two-sided chord Hilbert space $[\Delta_1\times\Delta_2]$.}
\label{fig:twosided}
\end{figure}
Next we investigate the property of the operator ${\sf D}$ when multiplied to other matter operators. In chord diagrams, products of matter operators behave as sources of parallel matter chords, and it would be natural to look for a kind of OPE formula which determines how to combine parallel matter chords into a single matter chord. A useful framework to study such a problem is the two-sided state formalism developed in \cite{Lin:2022rbf,Lin:2023trc}. To describe the product of $K$ matter operators of dimensions $\Delta_1,\cdots,\Delta_K$ in this formalism, one thinks of a half-disk with the $K$ matter chords extending inwards from its boundary semicircle without crossing. The semicircle is thus divided into $K+1$ segments by the endpoints of the $K$ matter chords. One may add arbitrary number of $H$-chords: say $n_j$ $H$-chords to the $j$-th boundary segment. An example is shown in Figure \ref{fig:twosided}. Then, to each such configuration of a half-disk with $K+\sum_jn_j$ parallel chords, one associates a basis vector $|n_0,n_1,\cdots,n_K\rangle$ of the two-sided chord Hilbert space $[\Delta_1\times\cdots\times\Delta_K]$. The product of matter operators ${\sf M}_\mini{\Delta_1},{\sf M}_\mini{\Delta_2},\cdots,{\sf M}_\mini{\Delta_K}$ can then be studied by matching the vector spaces $[\Delta_1\times\cdots\times\Delta_K]$ and $\oplus_j[\Delta_j]$ for suitable $\{\Delta_j\}$.

The two-sided chord Hilbert space admits the action of the chord Hamiltonians ${\sf H}_\mini{\text{L}},{\sf H}_\mini{\text{R}}$ acting on the left or the right ends of the boundary semicircle. They are expressed in terms of chord creation and annihilation operators as follows:
\begin{equation}
 {\sf H}_\mini{\text{L}}={\sf a}_\mini{\text{L}}+{\sf a}^\dagger_\mini{\text{L}},\qquad {\sf H}_\mini{\text{R}}={\sf a}_\mini{\text{R}}+{\sf a}^\dagger_\mini{\text{R}}.
\end{equation}
When multiplied onto the state $|n_0,\cdots,n_K\rangle\in[\Delta_1\times\cdots\times\Delta_K]$, the creation operators ${\sf a}^\dagger_\mini{\text{L}}$ or ${\sf a}^\dagger_\mini{\text{R}}$ add one $H$-chord to the leftmost or the rightmost segments:
\begin{align}
 {\sf a}^\dagger_\mini{\text{L}}|n_0,n_1,\cdots,n_K\rangle&\;=\; |n_0+1,n_1,\cdots,n_K\rangle,\nonumber \\
 {\sf a}^\dagger_\mini{\text{R}}|n_0,n_1,\cdots,n_K\rangle&\;=\; |n_0,n_1,\cdots,n_K+1\rangle.
\end{align}
On the other hand, the annihilation operators ${\sf a}_\mini{\text{L}}$ or ${\sf a}_\mini{\text{R}}$ absorb any one of the $(n_0+\cdots+n_K)$ $H$-chords. Taking into account that an $H$-chord may have to cross other $H$- or $M$-chords before reaching the boundary to get absorbed, one finds
\begin{align}
{\sf a}_\mini{\text{L}}|n_0,n_1,\cdots,n_K\rangle
\;=\;& [n_0]_q|n_0-1,n_1,\cdots,n_K\rangle+q^{n_0+\Delta_1}[n_1]_q|n_0,n_1-1,\cdots,n_K\rangle + ~\nonumber \\ &\cdots+
q^{(n_0+\cdots+n_{K-1})+(\Delta_1+\cdots+\Delta_K)}[n_K]_q|n_0,n_1,\cdots,n_K-1\rangle,
\end{align}
and a similar formula for the action of ${\sf a}_\mini{\text{R}}$. Each of the pairs ${\sf a}_\mini{\text{L}},{\sf a}^\dagger_\mini{\text{L}}$ and ${\sf a}_\mini{\text{R}},{\sf a}^\dagger_\mini{\text{R}}$ obeys the Arik-Coon $q$-oscillator algebra (\ref{ArikCoon}). One also has
\begin{align}
[{\sf H}_\mini{\text{L}},{\sf H}_\mini{\text{R}}]\;=\;
[{\sf a}_\mini{\text{L}},{\sf a}_\mini{\text{R}}]\;=\;
[{\sf a}^\dagger_\mini{\text{L}},{\sf a}^\dagger_\mini{\text{R}}]\;=\;&0,
\nonumber \\
[{\sf a}_\mini{\text{L}},{\sf a}^\dagger_\mini{\text{R}}]\;=\;
[{\sf a}_\mini{\text{R}},{\sf a}^\dagger_\mini{\text{L}}]\;\equiv\;&
q^{{\sf N}_\text{tot}},
\end{align}
where ${\sf N}_\text{tot}$ counts the total number of chords:
\begin{align}
q^{{\sf N}_\text{tot}}|n_0,\cdots,n_K\rangle
\;=\;& q^{(n_0+\cdots+n_K)+(\Delta_1+\cdots+\Delta_K)}|n_0,\cdots,n_K\rangle\,.
\end{align}

\paragraph{Representation-theoretic analysis}

When studying the isomorphisms between the Hilbert spaces $[\Delta_1\times\cdots\times\Delta_K]$ and $\oplus_j[\Delta_j]$, we restrict to those which commute with the action of the $q$-oscillator algebra of ${\sf a}_\mini{\text{L}},{\sf a}^\dagger_\mini{\text{L}},{\sf a}_\mini{\text{R}},{\sf a}^\dagger_\mini{\text{R}}$. A useful notion in studying such isomorphisms is the decomposition of Hilbert spaces into irreducible representations of the $q$-oscillator algebra. The relevant irreducible representations are those in which the spectrum of ${\sf N}_\text{tot}$ is bounded from below. In what follows we call the total number of $H$-chords $(n_0+\cdots+n_K)$ the {\em level}. The states annihilated by both ${\sf a}_\mini{\text{L}}$ and ${\sf a}_\mini{\text{R}}$ are called {\em primary} and play the role of the lowest weight states. The states obtained by multiplying some ${\sf a}^\dagger_\mini{\text{L}}$ or ${\sf a}^\dagger_\mini{\text{R}}$ on some primary states are called {\em descendants}. A lowest weight representation space called {\it Verma module} is spanned by a primary state and all its descendants.

The space $[\Delta_1\times\cdots\times\Delta_K]$ has a unique state at level zero, which is primary. The irreducible decomposition of this space amounts to finding all other primary states that may exist at nonzero level.

Let us begin with the case $K=1$. The space $[\Delta]$ is spanned by the primary state $|0,0\rangle$ and its descendants $|n_0,n_1\rangle={\sf a}_\mini{\text{L}}^{\dagger n_0}{\sf a}_\mini{\text{R}}^{\dagger n_1}|0,0\rangle$, so it is a single Verma module. The number of independent states in $[\Delta]$ at level $\ell$ is
\[
 d_1(\ell)\;=\;\ell+1. 
\]
Let us find out if the space has other primaries at nonzero level $\ell>0$ for special values of $\Delta$. Such states will be called {\it null states} because they are orthogonal to the whole Verma module with respect to the inner product which is defined from chord diagrams. Assuming the genericity of $q$, i.e. it is not a root of unity, ${\sf a}_\mini{\text{L}}$ and ${\sf a}_\mini{\text{R}}$ both define surjective linear maps from the subspace of level-$\ell$ states to that of level-$(\ell-1)$ states. These maps therefore both have one-dimensional kernel, and there will be one null state at level $\ell$ if the two kernels happen to coincide. Let the candidate null state be given by
\begin{equation}
 |\chi\rangle \;=\; \sum_{j=0}^\ell |j,\ell-j\rangle C_j
\end{equation}
with the coefficients $C_j$ to be determined. Then the condition ${\sf a}_\mini{\text{L}}|\chi\rangle={\sf a}_\mini{\text{R}}|\chi\rangle=0$ yields
\begin{equation}
\left(\begin{array}{ll}
\,1 & q^{\Delta+j}\\[1mm]
\,q^{\Delta+\ell-j-1} & 1\end{array}\right)
\left(\begin{array}{l}\,[j+1]_qC_{j+1}\\[1mm]\,[\ell-j]_qC_j\end{array}\right)\;=\;0\qquad(j=0,1,\cdots,\ell-1).
\end{equation}
Interestingly, for $\Delta=\frac{1-\ell}2$ the $2\times2$ matrix on the LHS has zero determinant for all $j$ and the above set of equations has a nontrivial solution:
\begin{equation}
 C_j\;=\; (-1)^jq^{-\frac{j(\ell-j)}2}\frac{[\ell]_q!}{[j]_q![\ell-j]_q!}.
\end{equation}
As the simplest nontrivial example, for $\Delta=-\frac12$ there is a null state at level $\ell=2$:
\begin{equation}
 |\chi\rangle \;=\; |2,0\rangle - (q^{\frac12}+q^{-\frac12})|1,1\rangle+|0,2\rangle\;\in\;[\Delta].
\label{nullvec2}
\end{equation}
In fact, not only the state $|\chi\rangle$ but also all its descendants are orthogonal to the whole Verma module. Therefore, for $\Delta=\frac{1-\ell}2$ an irreducible representation can be obtained by regarding $|\chi\rangle$ and all its descendants as zero. Such a representation is called a {\it degenerate representation}.

The form of the above null state is somewhat similar to that of the operator $\chi({\sf D})$ (\ref{nullvec}). This similarity can be made into a precise relation by introducing the notion of normal ordered products (of operators in the original one-sided formalism). We define it as in ordinary quantum field theories by subtracting contractions. For example,
\begin{align}
:\!{\sf H}{\sf H}\!: &\;=\; {\sf H}{\sf H}-\contraction{}{{\sf H}}{}{{\sf H}}{{\sf H}{\sf H}}\;=\; {\sf HH}-1,\nonumber \\
:\!{\sf H}{\sf M}_\mini{\Delta}{\sf H}\!: &\;=\; {\sf H}{\sf M}_\mini{\Delta}{\sf H}-\contraction{}{{\sf H}}{{\sf M}_\mini{\Delta}}{{\sf H}}{{\sf H}{\sf M}_\mini{\Delta}{\sf H}}\;=\; {\sf H}{\sf M}_\mini{\Delta}{\sf H}-q^\Delta{\sf M}_\mini{\Delta},\qquad\text{etc.}
\end{align}
Using this rule one can express the operator $\chi({\sf D})$ (\ref{nullvec}) as a normal ordered product
\begin{equation}
\chi({\sf D})\;=\; :\!{\sf H}^2{\sf D}\!:-(q^{\frac12}+q^{-\frac12}):\!{\sf H}{\sf D}{\sf H}\!:+:\!{\sf D}{\sf H}^2 \!:.
\end{equation}
The relation between $\chi({\sf D})$ and the state $|\chi\rangle$ (\ref{nullvec2}) has now become fully precise. The above result also leads us to expect that the null vector for ${\sf M}_{\Delta=\frac{1-\ell}2}$ can be expressed using normal ordered products as well.

Let us next turn to the case $K=2$. We first study the space $[\Delta_1\times\Delta_2]$ for generic $\Delta_1,\Delta_2$. Since the dimension of the subspace of level-$\ell$ states equals
\[
 d_2(\ell)\;=\;\frac{(\ell+1)(\ell+2)}2\;=\; d_1(\ell)+d_1(\ell-1)+\cdots+d_1(0),
\]
there is one primary state at each level $\ell$. The operator corresponding to the primary state at level $\ell$ may be regarded as a matter operator of dimension $\Delta_1+\Delta_2+\ell$. So we have the fusion rule
\[
 [\Delta_1\times\Delta_2]\;=\; \bigoplus_{\ell=0}^\infty[\Delta_1+\Delta_2+\ell].
\]

For special values of $\Delta_1,\Delta_2$, some primary states at higher levels may have to be regarded as null states. As an example, let us set $\Delta_1=-\frac12$ while keeping $\Delta_2=\Delta$ generic. Then at each of the levels $\ell=0,1,2$ there is one primary state. They take the form
\begin{align}
|({\sf D}{\sf M}_\mini{\Delta})_0\rangle&\;\equiv\; |0,0,0\rangle,\nonumber \\
|({\sf D}{\sf M}_\mini{\Delta})_1\rangle&\;\equiv\; f_\Delta(q)\left\{
\,-\,
(q^\Delta-q^{-\Delta})|1,0,0\rangle
\,+\,
(q^{\Delta-\frac12}-q^{-\Delta+\frac12})|0,1,0\rangle
\,+\,
(q^\frac12-q^{-\frac12})|0,0,1\rangle
\right\},\nonumber \\
|({\sf D}{\sf M}_\mini{\Delta})_2\rangle&\;\equiv\; |2,0,0\rangle - (q^{\frac12}+q^{-\frac12})|1,1,0\rangle + |0,2,0\rangle.
\label{null2}
\end{align}
The normalization factor $f_\Delta(q)$ for $|({\sf D}{\sf M}_\mini{\Delta})_1\rangle$ will be determined later. The state $|({\sf D}{\sf M}_\mini{\Delta})_2\rangle$ should be regarded as null since it clearly corresponds to the operator $\chi({\sf D}){\sf M}_\mini{\Delta}$. In fact, the state $|({\sf D}{\sf M}_\mini{\Delta})_2\rangle$ as well as all its descendants are null states. One can list up the basis states at level $\ell$ in the space $[(-\frac12)\times\Delta]$ and its null subspace by using the correspondence with the normal ordered operators:
\begin{alignat}{3}
&\text{states}\quad|n_0,n_1,n_2\rangle &&\quad\leftrightarrow\quad
:\!{\sf H}^{n_0}{\sf D}{\sf H}^{n_1}{\sf M}_\mini{\Delta}{\sf H}^{n_2}\!:
\quad&&(n_0+n_1+n_2=\ell) \nonumber \\
&\text{null states} &&\quad\leftrightarrow\quad
:\!{\sf H}^{n_0}\chi({\sf D}){\sf H}^{n_1}{\sf M}_\mini{\Delta}{\sf H}^{n_2}\!:
\quad&&(n_0+n_1+n_2+2=\ell)
\end{alignat}
The number of independent states at level $\ell(\ge 2)$, after eliminating null states, is thus given by
\begin{equation}
 d_2(\ell)-d_2(\ell-2)\;=\;2\ell+1\;=\; d_1(\ell)+d_1(\ell-1).
\end{equation}
This implies that the fusion rule involving a matter operator of dimension $-\frac12$ is
\begin{equation}
 [(-\tfrac12)\times\Delta]\;=\; [\Delta-\tfrac12]\oplus[\Delta+\tfrac12].
\label{DMOPE}
\end{equation}

\paragraph{Operator product relations}

The above analysis does not only give an isomorphism between the Hilbert spaces, but also seems to imply that the composite operators
\begin{align}
({\sf D}{\sf M}_\mini{\Delta})_0&\;\equiv\;{\sf D}{\sf M}_\mini{\Delta},
\nonumber \\
({\sf D}{\sf M}_\mini{\Delta})_1&\;\equiv\;f_\Delta(q)\left\{
-(q^\Delta-q^{-\Delta}){\sf H}{\sf D}{\sf M}_\mini{\Delta}
+(q^{\Delta-\frac12}-q^{-\Delta+\frac12}){\sf D}{\sf H}{\sf M}_\mini{\Delta}
+(q^{\frac12}-q^{-\frac12}){\sf D}{\sf M}_\mini{\Delta}{\sf H}\right\}
\label{compop}
\end{align}
are equivalent respectively to matter operators of dimensions $\Delta-\frac12$ and $\Delta+\frac12$. The relations (\ref{DMOPE}) and (\ref{compop}) are analogous to the boundary Liouville OPE (\ref{bOPE}) involving the operator $B_{-\frac b2}$. It is interesting to find out to what extent these relations hold in correlators.

Let us begin by fixing the normalizations. The norm of any composite operator can naturally be determined from the two-point function of the operator and its conjugate. According to this definition, the squared norm of $|({\sf D}{\sf M}_\mini{\Delta})_0\rangle$ is given by the two-point function of ${\sf D}{\sf M}_\mini{\Delta}$ and its conjugate ${\sf M}_\mini{\Delta}{\sf D}$, which simply equals 1.
\begin{equation}
\langle({\sf D}{\sf M}_\mini{\Delta})_0|({\sf D}{\sf M}_\mini{\Delta})_0\rangle\;=\;
\langle 0|
\contraction[1.3ex]{}{{\sf M}}{{}_\mini{\Delta}{\sf D}{\sf D}}{{\sf M}}
{\sf M}_\mini{\Delta}\contraction[.7ex]{}{{\sf D}}{}{{\sf D}}{\sf D}{\sf D}{\sf M}_\mini{\Delta}
|0\rangle\;=\; 1.
\end{equation}
So we normalize $|({\sf D}{\sf M}_\mini{\Delta})_0\rangle$ to simply correspond to ${\sf D}{\sf M}_\mini{\Delta}$. The computation of the squared norm of $|({\sf D}{\sf M}_\mini{\Delta})_1\rangle$ is a bit more cumbersome and involves summation over some chord diagrams. The result can be used to determine $f_\Delta(q)$ as follows:
\begin{align}
\langle({\sf D}{\sf M}_\mini{\Delta})_1|({\sf D}{\sf M}_\mini{\Delta})_1\rangle&\;=\;f_\Delta(q)^2\cdot
(q^{\frac12}-q^{-\frac12})(q^{\Delta}-q^{-\Delta})(q^{\Delta-\frac12}-q^{-\Delta+\frac12}),\nonumber \\
\therefore\qquad f_\Delta(q)&\;\equiv\;\left\{(q^{\frac12}-q^{-\frac12})(q^{\Delta}-q^{-\Delta})(q^{\Delta-\frac12}-q^{-\Delta+\frac12})\right\}^{-\frac12}.
\end{align}

In order to test the equivalence of the composite operators $({\sf DM}_\mini{\Delta})_{0,1}$ to ${\sf M}_\mini{\Delta\mp\frac12}$, let us now compute their thermal two-point correlators. First, the correlator of $({\sf D}{\sf M}_\mini{\Delta})_0$ can be expressed as
\begin{equation}
 \langle 0|\rme^{-\beta_1{\sf H}}
\contraction{}{({\sf D}{\sf M}_\mini{\Delta})_0}{\rme^{-\beta_2{\sf H}}}{({\sf D}{\sf M})_0^\dagger}
({\sf D}{\sf M}_\mini{\Delta})_0\,\rme^{-\beta_2{\sf H}}({\sf D}{\sf M}_\mini{\Delta})_0^\dagger|0\rangle\;=\;
\int_0^\pi\prod_{i=1}^2\frac{\mathrm d\theta_i\mu(\theta_i)\rme^{-\beta_iE(\theta_i)}}{2\pi}
\langle\theta_1|
\contraction{}{({\sf D}{\sf M}_\mini{\Delta})_0}{|\theta_2\rangle\langle\theta_2|}{({\sf D}{\sf M}_\mini{\Delta})_0^\dagger}
({\sf D}{\sf M}_\mini{\Delta})_0|\theta_2\rangle\langle\theta_2|({\sf D}{\sf M}_\mini{\Delta})_0^\dagger|0\rangle\,.
\end{equation}
The integrand can be rewritten further by inserting ${\sf H}$-eigenstates between ${\sf D}$ and ${\sf M}_\mini{\Delta}$:
\begin{align}
\langle\theta_1|
\contraction{}{({\sf D}{\sf M}_\mini{\Delta})_0}{|\theta_2\rangle\langle\theta_2|}{({\sf D}{\sf M}_\mini{\Delta})_0^\dagger}
({\sf D}{\sf M}_\mini{\Delta})_0|\theta_2\rangle\langle\theta_2|({\sf D}{\sf M}_\mini{\Delta})_0^\dagger|0\rangle&\;=\;
\int\frac{\mathrm d\theta\mu(\theta)}{2\pi}
\frac{\mathrm d\theta'\mu(\theta')}{2\pi}\langle\theta_1|
\contraction[1.3ex]{}
{{\sf D}}{|\theta\rangle\langle\theta|
{\sf M}_\mini{\Delta}|\theta_2\rangle\langle\theta_2|{\sf M}_\mini{\Delta}
|\theta'\rangle\langle\theta'|}{{\sf D}}
{\sf D}|\theta\rangle\langle\theta|
\contraction[.7ex]{}{{\sf M}}{{}_\mini{\Delta}|\theta_2\rangle\langle\theta_2|}{{\sf M}}
{\sf M}_\mini{\Delta}|\theta_2\rangle\langle\theta_2|{\sf M}_\mini{\Delta}
|\theta'\rangle\langle\theta'|{\sf D}
|0\rangle
\nonumber \\ &\;=\;
\int\frac{\mathrm d\theta\mu(\theta)}{2\pi}G(-\tfrac12|\theta_1,\theta)G(\Delta|\theta,\theta_2)
\nonumber \\ &\;=\;
\sum_{\pm}\frac{G(\Delta|\theta_1\pm\frac{\rmi\lambda}2,\theta_2)}{1-\rme^{\pm2\rmi\theta_1}}\;=\; G(\Delta-\tfrac12|\theta_1,\theta_2),
\label{24rel1}
\end{align}
where we used (\ref{G-1/2}) at the third equality. This supports our claim that ${\sf DM}_\mini{\Delta}$ is equivalent to a matter operator of dimension $\Delta-\frac12$. For $({\sf DM}_\mini{\Delta})_1$, we first work out its matrix element between ${\sf H}$-eigenstates:
\begin{align}
\langle\theta_1|({\sf D}{\sf M}_\mini{\Delta})_1|\theta_2\rangle&\;=\;
\int_0^\pi\frac{\mathrm d\theta\mu(\theta)}{2\pi}\langle\theta_1|{\sf D}|\theta\rangle\langle\theta|{\sf M}_\mini{\Delta}|\theta_2\rangle\cdot F_\Delta(\theta_1,\theta,\theta_2;q), \nonumber \\
F_\Delta(\theta_1,\theta,\theta_2;q) &\;=\;
f_\Delta(q)\left\{
-(q^\Delta-q^{-\Delta})E(\theta_1)
+(q^{\Delta-\frac12}-q^{-\Delta+\frac12})E(\theta)
+(q^\frac12-q^{-\frac12})E(\theta_2)
\right\}.
\end{align}
The way the coefficient $F_\Delta$ is defined is somewhat analogous to how the boundary Liouville OPE coefficient $c_-$ was defined in (\ref{c-2}). Moreover, $F_\Delta$ factorizes for $\theta=\theta_1\pm\frac{\rmi\lambda}2$ in the same way that $c_-$ does as we have seen in (\ref{c-3}).
\begin{equation}
 F_\Delta(\theta_1,\theta_1\pm\tfrac{\rmi\lambda}2,\theta_2;q)\;=\;
 f_\Delta(q)(1-q)^{\frac12}q^{-\Delta}\rme^{\pm \rmi\theta_1}
(1-\rme^{\mp\rmi\theta_1+\rmi\theta_2}q^{\Delta-\frac12})
(1-\rme^{\mp\rmi\theta_1-\rmi\theta_2}q^{\Delta-\frac12})\,.
\label{FDelta}
\end{equation}
Using this one can check that
\begin{align}
\langle\theta_1|
\contraction{}{({\sf D}{\sf M}_\mini{\Delta})_1}{|\theta_2\rangle\langle\theta_2|}{({\sf D}{\sf M}_\mini{\Delta})_1^\dagger}
({\sf D}{\sf M}_\mini{\Delta})_1|\theta_2\rangle\langle\theta_2|({\sf D}{\sf M}_\mini{\Delta})_1^\dagger|0\rangle&\;=\;
\sum_{\pm}\frac{G(\Delta|\theta_1\pm\frac{\rmi\lambda}2,\theta_2)}{1-\rme^{\pm2\rmi\theta_1}}F_\Delta(\theta_1,\theta_1\pm\tfrac{\rmi\lambda}2,\theta_2;q)^2
\nonumber \\&\;=\; G(\Delta+\tfrac12|\theta_1,\theta_2).
\label{24rel2}
\end{align}
This supports our claim that $({\sf DM}_\mini{\Delta})_1$ is equivalent to a matter operator of dimension $\Delta+\frac12$.

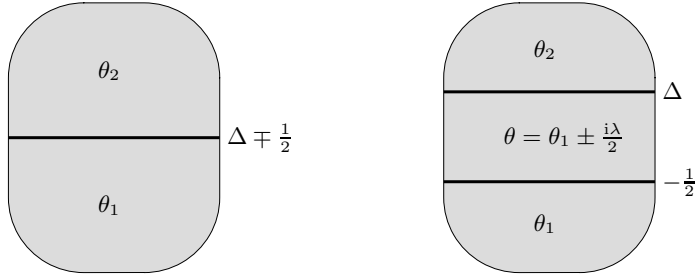
\begin{figure}
\begin{center}
\begin{tabular}{ccc}
\begin{overpic}[scale=1]{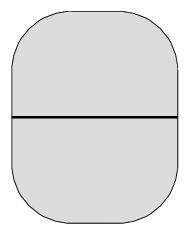}
\put(78,48){\footnotesize $\Delta\mp\frac12$}
\put(35,70){\footnotesize $\theta_2$}
\put(35,25){\footnotesize $\theta_1$}
\end{overpic}
 &~\hskip15mm~ &
\begin{overpic}[scale=1]{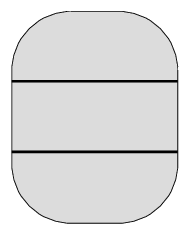}
\put(78,33){\footnotesize $-\frac12$}
\put(78,63){\footnotesize $\Delta$}
\put(35,77){\footnotesize $\theta_2$}
\put(35,19){\footnotesize $\theta_1$}
\put(25,48){\footnotesize $\theta=\theta_1\pm\frac{\rmi\lambda}2$}
\end{overpic}
\end{tabular} 
\caption{The two and four-point functions that are related to each other.}\label{fig:24pts}
\end{center} 
\end{figure}

\paragraph{Recursion relations for $G(\Delta|\theta_1,\theta_2)$}

The equations (\ref{24rel1}) and (\ref{24rel2}) relates the two-point function $G(\Delta\mp\frac12|\theta_1,\theta_2)$ (shown on the left of Figure \ref{fig:24pts}) to the uncrossed four-point function with $\theta=\theta_1\pm\frac{\rmi\lambda}2$ for the region between the two matter chords (the right). These correlators should all have an interpretation as the squared norm of some states in suitable two-sided chord Hilbert spaces. It would then be natural to think that the relations among correlators follow from some linear relations among those states. In fact, by following this line of arguments, one can derive a set of two-term recursion relations for $G(\Delta|\theta_1,\theta_2)$ that is powerful enough to reproduce (\ref{GDelta}).

We are interested in the four states represented by the following graphs:
\begin{equation}
|s_\pm\rangle\;=\;
\raisebox{-9mm}{
\begin{overpic}[scale=1]{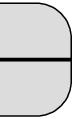}
\put(65,44){\footnotesize $\Delta\pm\frac12$}
\put(20,20){\footnotesize $\theta_1$}
\put(20,69){\footnotesize $\theta_2$}
\end{overpic}
}\hskip10mm,\qquad
|t_\pm\rangle\;=\;
\raisebox{-9mm}{
\begin{overpic}[scale=1]{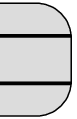}
\put(65,22){\footnotesize $-\frac12$}
\put(65,67){\footnotesize $\Delta$}
\put(20,10){\footnotesize $\theta_1$}
\put(20,45){\footnotesize $\theta$}
\put(20,79){\footnotesize $\theta_2$}
\end{overpic}
}\hskip5mm.
\quad\left(\theta=\theta_1\pm\tfrac{\rmi\lambda}2\right)
\end{equation}
The states $|s_\pm\rangle$ are the unique eigenstates of ${\sf H}_\mini{\text{L}},{\sf H}_\mini{\text{R}}$ with eigenvalues $E(\theta_1),E(\theta_2)$ in the respective Hilbert spaces $[\Delta\pm\frac12]$. Similarly, the states $|t_\pm\rangle\in[(-\frac12)\times\Delta]$ are both in the eigenspace of ${\sf H}_\mini{\text{L}},{\sf H}_\mini{\text{R}}$ with the eigenvalues $E(\theta_1),E(\theta_2)$. The isomorphism (\ref{DMOPE}) implies that this eigenspace is isomorphic to the space spanned by $|s_\pm\rangle$; in particular, it is two-dimensional. This agrees nicely with the fact that the energy $E(\theta)$ in the region between the ${\sf D}$-chord and ${\sf M}_\mini{\Delta}$-chord can take two distinct values. The states $|s_\pm\rangle$ and $|t_\pm\rangle$ should therefore be related by a linear transformation.

The bra states $\langle s_\pm|, \langle t_\pm|$ are represented by the left-right reflection of the above graphs for the kets. The states $|s_+\rangle$ and $|s_-\rangle$ are orthogonal to each other, and they have the squared-norm
\begin{equation}
 \langle s_\pm|s_\pm\rangle = G(\Delta\pm\tfrac12|\theta_1,\theta_2).
\end{equation}
Likewise, $|t_+\rangle$ and $|t_-\rangle$ are orthogonal to each other. Regarding their norm, we assume that $G(-\frac12|\theta_1,\theta)$ have delta-functional support at $\theta=\theta_1\pm\frac{\rmi\lambda}2$ but do not want to use its explicit form (\ref{G-1/2}). So we express $G(-\frac12|\theta_1,\theta)$ and $\langle t_\pm|t_\pm\rangle$ in terms of some unknown functions $g_\pm(\theta)$ as follows:
\begin{equation}
G(-\tfrac12|\theta_1,\theta)=\sum_{\pm}\frac{g_\pm(\theta_1)2\pi\delta(\theta_1\pm\frac{\rmi\lambda}2-\theta)}{\mu(\theta)},\qquad
\langle t_\pm|t_\pm\rangle \;=\; g_\pm(\theta_1)G(\Delta|\theta_1\pm\tfrac{\rmi\lambda}2,\theta_2).
\end{equation}
Our goal is thus to derive recursion relations that can determine $G(\Delta|\theta_1,\theta_2)$ and $g_\pm(\theta)$.

The relations (\ref{24rel1}), (\ref{24rel2}) can now be expressed in terms of the states $|s_\pm\rangle,|t_\pm\rangle$ as
\begin{equation}
 \langle s_-|s_-\rangle = \sum_\pm\langle t_\pm|t_\pm\rangle,\qquad
 \langle s_+|s_+\rangle = \sum_\pm\langle t_\pm|t_\pm\rangle F_\pm^2,
\end{equation}
where $F_\pm$ is a shorthand for $F_\Delta(\theta_1,\theta_1\pm\frac{\rmi\lambda}2,\theta_2;q)$ (\ref{FDelta}). If there is a linear transformation relating $|s_\pm\rangle$ and $|t_\pm\rangle$, it must take the following form (up to unphysical sign choices):
\begin{equation}
\begin{aligned}
& |s_-\rangle \;=\; |t_+\rangle + |t_-\rangle,\\
& |s_+\rangle \;=\; |t_+\rangle F_+ + |t_-\rangle F_-,
\end{aligned}
\qquad
 |t_\pm\rangle \;=\; \frac{|s_+\rangle- |s_-\rangle F_\mp}{F_\pm-F_\mp}.
\end{equation}
At this point we notice that the condition $\langle t_+|t_-\rangle=0$, when rewritten in terms of $|s_\pm\rangle$, turns into a nontrivial recursion relation for $G(\Delta|\theta_1,\theta_2)$:
\begin{align}
0&\;=\;\langle s_+|s_+\rangle + \langle s_-|s_-\rangle F_+F_-
\nonumber \\ &\;=\;
G(\Delta+\tfrac12|\theta_1,\theta_2)-\frac{G(\Delta-\tfrac12|\theta_1,\theta_2)}{(1-q^{2\Delta})(1-q^{2\Delta-1})}\prod_{\pm,\pm'}(1-q^{\Delta-\frac12}\rme^{\pm\rmi\theta_1\pm'\rmi\theta_2}),
\label{recG1}
\end{align}
which is solved by (\ref{GDelta}). Similarly, by expressing $\langle s_+|s_-\rangle=0$ in terms of $|t_\pm\rangle$ one obtains another recursion relation
\begin{equation}
\frac{g_+(\theta_1)G(\Delta|\theta_1+\frac{\rmi\lambda}2,\theta_2)}
{g_-(\theta_1)G(\Delta|\theta_1-\frac{\rmi\lambda}2,\theta_2)}
= -\rme^{-2\rmi\theta_1}\prod_\pm\frac{(1-q^{\Delta-\frac12}\rme^{+\rmi\theta_1\pm\rmi\theta_2})}{(1-q^{\Delta-\frac12}\rme^{-\rmi\theta_1\pm\rmi\theta_2})},
\label{recG2}
\end{equation}
which is again solved by (\ref{GDelta}) and its consequence $g_\pm(\theta)=(1-\rme^{\pm 2\rmi\theta})^{-1}$ (\ref{G-1/2}).

Some remarks are in order concerning the uniqueness of the solution. Recall that $G(\Delta|\theta_1,\theta_2)$ is periodic under $\theta_i\to\theta_i+2\pi$ in addition to obeying the recursion relation (\ref{recG2}) under $\theta_i\to\theta_i+\rmi\lambda$. Having two shift relations for a single complex variable, like (\ref{bGbS}) for the functions $\bG(x)$ and $\bS(x)$, is usually taken to be strong enough for a function to be determined unambiguously. As for the dependence of $G(\Delta|\theta_1,\theta_2)$ on $\Delta$, we have found only one shift relation (\ref{recG1}) under $\Delta\to\Delta+1$. But actually $G(\Delta|\theta_1,\theta_2)$ has to be periodic under $\Delta\to\Delta+\frac{2\pi\rmi}\lambda$ because so is its integral transform $G(\Delta|\beta_1,\beta_2)$ which is a power series in $q^\Delta=\rme^{-\lambda\Delta}$ according to diagrammatics.

\paragraph{Three-point functions and chord junctions}

One may naturally guess that the inner products of $|s_\pm\rangle$ and $|t_\pm\rangle$ would correspond to three-point functions, but upon a closer inspection one encounters a subtle problem.

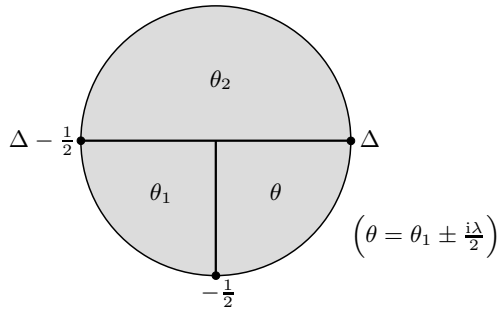
\begin{figure}[h]
\begin{center}
\begin{overpic}[scale=1]{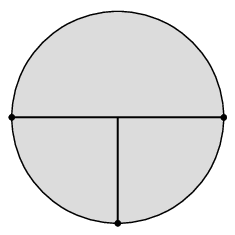}
\put(98,48){\footnotesize $\Delta$}
\put(45,-3){\footnotesize $-\frac12$}
\put(-19,48){\footnotesize $\Delta-\frac12$}
\put(28,30){\footnotesize $\theta_1$}
\put(68,30){\footnotesize $\theta$}
\put(48,68){\footnotesize $\theta_2$}
\put(95,16){\footnotesize $\Big(\theta=\theta_1\pm\frac{\rmi\lambda}2\Big)$}
\end{overpic}
\end{center}
\caption{An example of three-point functions.}\label{fig:3pt}
\end{figure}

As an example, let us consider the correlator corresponding to the diagram in Figure \ref{fig:3pt}. By moving the operator ${\sf D}$ (of weight $-\frac12$) towards ${\sf M}_\mini{\Delta}$ one obtains a diagram representing the inner product $\langle s_-|t_\pm\rangle$. On the other hand, by moving ${\sf D}$ towards ${\sf M}_\mini{\Delta-\frac12}$ one obtains a diagram for a different inner product $\langle t'_\mp|s'_+\rangle$, where $|s'_\pm\rangle$ and $|t'_\pm\rangle$ are related to $|s_\pm\rangle, |t_\pm\rangle$ by a change of parameters. Explicitly,
\begin{equation}
|s'_\mp\rangle\;=\;
\raisebox{-9mm}{
\begin{overpic}[scale=1]{1ch.eps}
\put(65,44){\footnotesize $\Delta'\mp\frac12$}
\put(20,20){\footnotesize $\theta$}
\put(20,69){\footnotesize $\theta_2$}
\end{overpic}
}\hskip10mm,\qquad
|t'_\mp\rangle\;=\;
\raisebox{-9mm}{
\begin{overpic}[scale=1]{2ch.eps}
\put(65,22){\footnotesize $-\frac12$}
\put(65,67){\footnotesize $\Delta'$}
\put(20,10){\footnotesize $\theta$}
\put(20,45){\footnotesize $\theta_1$}
\put(20,79){\footnotesize $\theta_2$}
\end{overpic}
}\hskip5mm.
\qquad\left(\;\begin{aligned}
\Delta'&=\Delta-\tfrac12 \\[1mm]
\theta_1&=\theta\mp\tfrac{\rmi\lambda}2\end{aligned}\;\right)
\end{equation}
By computing the two inner products explicitly one finds
\begin{align}
\langle s_-|t_\pm\rangle
=
\frac{G(\Delta|\theta,\theta_2)}{1-\rme^{\pm 2\rmi\theta_1}},
\qquad
\langle t'_\mp|s'_+\rangle
=
\frac{G(\Delta|\theta,\theta_2)}{1-\rme^{\mp 2\rmi\theta}}
\cdot\rmi\rme^{\mp\rmi\theta}\sqrt{\frac{1-q^{2\Delta-1}}{1-q^{2\Delta-2}}}
\label{stprod}
\end{align}
for $\theta=\theta_1\pm\frac{\rmi\lambda}2$. There is a nontrivial mismatch between the two, though a part of it can be attributed to the difference in the integration measure:
\begin{equation}
\frac{\mathrm d\theta_1}{2\pi}
\frac{\mu(\theta_1)}{1-\rme^{\pm 2\rmi\theta_1}}
\;=\;
\frac{\mathrm d\theta}{2\pi}
\frac{\mu(\theta)}{1-\rme^{\mp 2\rmi\theta}}.
\end{equation}

Hopefully this mismatch is not indicating a breakdown of our argument and can be resolved by a more systematic analysis. For the moment, let us just point out that it looks reminiscent of the following well-known fact: the Clebsch-Gordan coefficient $\langle\ell_1m_1,\ell_2m_2|\ell_3m_3\rangle$ is not equal to $\langle\ell_1m_1|\ell_2(-m_2),\ell_3m_3\rangle$. Instead, the Wigner's 3$j$ symbol
\begin{equation}
\left(\begin{array}{ccc}\ell_1&\ell_2&\ell_3\\[-1mm]m_1&m_2&m_3\end{array}\right)
\equiv\frac{(-1)^{\ell_1-\ell_2-m_3}}{\sqrt{2\ell_3+1}}
\langle\ell_1m_1,\ell_2m_2|\ell_3(-m_3)\rangle
\end{equation}
is known to be invariant (up to sign) under permutations of the three spins. Perhaps the three-point function in question, which should be symmetric in the three operators, will be related to $\langle s_-|t_\pm\rangle$ and $\langle t'_\mp|s'_+\rangle$ in the same way that the $3j$ symbol is related to the Clebsch-Gordan coefficient.

\section{Concluding remarks}\label{sec:concl}

In this paper we have discussed a number of interesting properties of degenerate matter operators in DSSYK model, focusing particularly on the most basic one ${\sf D}={\sf M}_\mini{-\frac12}$. Perhaps they are powerful enough that one can use them to reproduce all the exact results that were derived using diagrammatics. We were just able to derive the recursion relations for the two-point function $G(\Delta|\theta_1,\theta_2)$. It would be interesting to consider how our argument can be generalized to higher-point and crossed correlation functions.

The special properties of degenerate operators are encoded in the null vector equation. For the most basic degenerate operator ${\sf D}$, we have found using diagrammatics and representation-theoretic analysis that the null vector takes the form (\ref{nullvec}) or (\ref{nullvec2}). There are other degenerate operators of dimension $\Delta=\frac{1-\ell}2$ which have a null vector at level $\ell$. We expect that they have properties similar to $B_{(1-\ell)b/2}$ in Liouville CFT and correspond to the $\ell$-dimensional representation of ${\cal U}_{q^{1/2}}(\mathfrak{su}_{1,1})$.

We also obtained some incomplete results which suggest that the matter chords may form junctions. This might look somewhat surprising if one recalls that the disorder average was taken over the random couplings defining matter operators, as it would only allow the matter operators of the same species to be paired. Even if matter chords are really allowed to form junctions, it remains for us to understand how to translate between the inner products of two-sided states and general correlators.

Although we have successfully imported many algebraic techniques from boundary Liouville CFT to DSSYK model, the correspondences of physical quantities between the two theories were not very impressive. This is because the quantum group that underlies the Liouville CFT is in fact two copies of quantum groups that are related by $b\leftrightarrow\frac1b$ duality \cite{Faddeev:1999fe}. On the other hand, the quantum group behind the DSSYK model is a single copy of ${\cal U}_{q^{1/2}}(\mathfrak{su}_{1,1})$, so neither the two-point function $G(\Delta|\theta_1,\theta_2)$ nor its recursion relation exhibit any sort of duality. 

Finally, let us note that there is an interesting triality between the DSSYK model, 4d ${\cal N}=2$ pure $\mathrm{SU}(2)$ SYM and the quantization of $\mathrm{SL}(2,\mathbb{C})$ Chern-Simons theory on a sphere with two irregular singularities \cite{Gaiotto:2024kze}. Under this triality, the algebra of ${\sf H}$ and degenerate matter operators in DSSYM maps to the ``K-theoretic Coulomb branch algebra'' of line operators in the SYM and the ``Skein algebra'' of space-like (open and closed) Wilson line defects in the CS theory. Indeed, our null vector equation (\ref{nullvec}) follows from the operator product relation of the fundamental Wilson loop $w_1$ and dyonic loops $T_a~(a\in\mathbb{Z})$ in the SYM:
\begin{equation}
 w_1T_a = q^{-\frac14} T_{a+1}+q^{\frac14}T_{a-1},\qquad
 T_aw_1 = q^{\frac14} T_{a+1}+q^{-\frac14}T_{a-1}
\end{equation}
with the identification ${\sf H}=(1-q)^{-\frac12}w_1$ and ${\sf D}=T_0$. The triality may give a useful insight into the algebraic aspects of the DSSYK model, in particular when studying the properties of matter chords of generic dimensions and their junctions.

\paragraph{Disclaimer}

We certify that the opinions expressed herein are only those of the authors. They do not represent the official views of Japan Ministry of Defense or National Defense Academy.

\newpage

\providecommand{\href}[2]{#2}

\end{document}